\documentclass{aa}  
\newcommand{\citeg}[1]{\citep[e.g.,][]{#1}}

\usepackage{graphicx}
\usepackage{txfonts}
\usepackage{amsmath}	
\usepackage{wasysym}
\usepackage{xspace}
\usepackage{float}
\usepackage[normalem]{ulem}
\usepackage{wrapfig}
\usepackage{xcolor}
\usepackage{hyperref}

\begin{document} 

\title{The galaxy mass-size relation in CARLA clusters and proto-clusters at $1.4<z<2.8$: larger cluster galaxy sizes}



\author{Anton V. Afanasiev
    \inst{1}\thanks{anton.afanasiev@obspm.fr}
    \and Simona Mei \inst{1}\fnmsep\inst{2}
  \and Hao Fu \inst{3}
   \and Francesco Shankar \inst{3}
   \and Stefania Amodeo \inst{4}
   \and Daniel Stern \inst{2}
    \and Elizabeth A. Cooke \inst{5}
    \and Anthony H. Gonzalez \inst{6}
      \and Gaël Noirot \inst{7}
     \and Alessandro Rettura \inst{2}
      \and Dominika Wylezalek \inst{8} 
    \and Carlos De Breuck \inst{9}
   \and Nina A. Hatch \inst{10}
    \and Spencer A. Stanford \inst{11}
   \and Joël Vernet \inst{9}
}

\institute{Université Paris Cité, CNRS, Astroparticule et Cosmologie, F-75013 Paris, France
\and Jet Propulsion Laboratory, Cahill Center for Astronomy \& Astrophysics, California Institute of Technology, 4800 Oak Grove Drive, Pasadena, California, USA
\and School of Physics \& Astronomy, University of Southampton, Highfield, Southampton SO17 1BJ, UK
\and Universit\'{e} de Strasbourg, CNRS, Observatoire astronomique de Strasbourg, UMR 7550, F-67000 Strasbourg, France
\and National Physical Laboratory, Hampton Road, Teddington, Middlesex, TW11 0LW, UK
\and Department of Astronomy, University of Florida, Gainesville, FL 32611-2055, USA
\and Department of Astronomy \& Physics, Saint Mary’s University, 923 Robie Street, Halifax, NS B3H 3C3, Canada
\and Zentrum für Astronomie der Universität Heidelberg, Astronomisches Rechen-Institut, Mönchhofstr 12-14, D-69120 Heidelberg, Germany
\and European Southern Observatory, Karl-Schwarzschildstrasse 2, 85748 Garching, Germany
\and School of Physics and Astronomy, University of Nottingham, University Park, Nottingham NG7 2RD, UK
\and Department of Physics, University of California, One Shields Avenue, Davis, CA 95616, USA
}  
\titlerunning{The Galaxy Mass-Size Relation in CARLA}
 \date{Received July 29, 2022; accepted November 21, 2022}  
  
 
  \abstract{
  We study the galaxy mass-size relation in 15 spectroscopically confirmed clusters at $1.4<z<2.8$ from the CARLA survey. Our clusters span a total stellar mass in the range $11.3<\log(M^c_*/M_{\astrosun})<12.6$ (with an approximate halo mass in the range $13.5 \lesssim \log(M^c_h/M_{\astrosun}) \lesssim 14.5$). Our main finding is that cluster passive early-type galaxies (ETGs) at $z \gtrsim 1.5$ with a mass ${\rm log}(M/M_{\astrosun})>10.5$ are systematically $\gtrsim 0.2-0.3 {\rm dex}$ larger ($\gtrsim 3\sigma$) than field ETGs at a similar redshift and mass from the CANDELS survey. The passive ETG average size evolution with redshift is slower at $1<z<2$ when compared to the field. This could be explained by early-epoch differences in the formation and early evolution of galaxies in haloes of a different mass, as predicted by models. It does not exclude that other physical mechanisms, such as strong compaction and gas dissipation in field galaxies, followed by a sequence of mergers may have also played a significant role in the field ETG evolution, but not necessarily in the evolution of cluster galaxies. Our passive ETG mass-size relation shows a tendency to flatten at $9.6<{\rm log}(M/M_{\astrosun})<10.5$, where the average size is $\mathrm{log}(R_e/\mathrm{kpc}) = 0.05 \pm 0.22$, which is broadly consistent with galaxy sizes in the field and in the local Universe. This implies that galaxies in the low end of the mass-size relation do not evolve much from $z \sim 2$ to the present, and that their sizes evolve in a similar way in clusters and in the field. Brightest cluster galaxies lie on the same mass-size relation as satellites, suggesting that their size evolution is not different from satellites at redshift z $\gtrsim$ 2. Half of the active early-type galaxies, which are $30\%$ of our ETG sample, follow the field passive galaxy mass-size relation, and the other half follow the field active galaxy mass-size relation. These galaxies likely went through a recent merger or neighbor galaxy interaction, and would most probably quench at a later epoch and increase the fraction of passive ETGs in clusters. We do not observe a large population of compact galaxies (only one), as is observed in the field at these redshifts, implying that the galaxies in our clusters are not observed in an epoch close to their compaction.
  }
  


   \maketitle

\section{Introduction} \label{sec:intro}
In the local Universe and up to $z=3$, the most massive galaxies are also among the largest \citep{Kaufmann03,Gadotti09,Poggianti13a,Huertas13a,FernandezLorenzo13,Delaye14,Belli14,vanDokkum15}. For example, local elliptical galaxies follow a rather tight relation with an intrinsic scatter of less than 0.3~dex \citep{Nair11,Bernardi11a,Bernardi11b,Bernardi14}. 
This dependence is called the galaxy mass-size relation (MSR), and provides insight into the past and present evolution of galaxies. 

The first results on the mass-size and size-luminosity relation at $z=1$ and beyond were reported by \citet{Trujillo04,Trujillo06} and \citet{McIntosh05}, who initially did not find strong size differences for massive ($M_*>2-3\times10^{10}~M_{\astrosun}$) galaxies at $2<z<3$ compared to $z=0$, but they later reported that galaxies at $z=2.5$ are two times smaller on average. Later results \citep{Trujillo11,Mosleh11,Dutton11,Szomoru12} confirm that the stellar MSR was already in place at $z=1$, but its normalization increased at low redshift.

The arrival of the new data, such as the Cosmic Assembly Near-infrared Deep Extragalactic Legacy Survey (CANDELS; PI: S. Faber, H. Ferguson; \citealp{koekemoer11}; \citealp{grogin11}), and new observational techniques, such as strong lensing \citep{Yang21}, confirm this view. \citet{vanderWel12,vanderWel14} measure the MSR redshift evolution for both passive and star-forming galaxies in the field in the redshift range $0<z<3$. They demonstrate that the slope of the MSR does not evolve for either population. However, galaxies become more compact with increasing redshift, which is explained by \citet{Carollo13} by the fact that the Universe was more dense in earlier times, and the galaxy density evolves approximately as the density of the Universe. \citet{Dimauro19} analyze the MSR of bulges and disks, and find that they follow different MSRs. Their MSR weakly depends on the morphology of the host galaxy, and the sizes of disks do not depend on their star-formation activity. They conclude that quenching did not affect disk structures.

The shape of the MSR is consistent with a scenario in which galaxy growth is dominated by star formation due to cold gas accretion up to a certain mass (which is redshift- and size-dependent, corresponding to $M=10^{11}M_{\astrosun}$ at $z=2$ and $R_e=1$~kpc approximately) and by galaxy mergers at higher masses \citeg{Shankar13,vanDokkum15,Zanisi20}
In fact, hierarchical models could explain the fast size growth of giant elliptical galaxies only by sequential minor dry mergers since $z=2$ \citep{Naab09,Trujillo11, Newman12,vanDokkum15}.
On the other hand, spiral galaxies do not require minor mergers since their growth can be attributed to cold gas accretion \citep{Dekel09}.

It is less clear though if galaxies in clusters and in the field evolved in the same way. In the local Universe, semi-analytical models predict a moderate to strong environmental dependence \citep{Shankar14a}; however most of the observational results agree that this relation is independent of environment \citeg{Guo09,Weinmann09,Cappellari13,Huertas13b,Mosleh18}. \citet{Poggianti13a} find that cluster early-type galaxies (ETGs) are smaller than those in the field; however, they included a large fraction of S0 galaxies which appear to have smaller radii than elliptical galaxies at a fixed stellar mass \citep{Bernardi13,Huertas13a} and different environmental relations \citep{Erwin12,Silchenko18}. In another work, \citet{Huang18} find that massive galaxies in clusters are as much as 20\%-40\% larger than in the field based on deep observations with the Hyper Suprime-Cam (see also \citealp{Yoon17}). For spiral galaxies, the environmental dependence of the MSR is more pronounced: its scatter is much larger \citep{Maltby10,Cappellari13,Lange15} and disks are smaller in clusters \citep{Kuchner17,Demers19}. This means that dense environments either destroy disks or inhibit their growth, for example through tidal interactions, ram-pressure, and/or strangulation \citep{2006PASP..118..517B}.


At intermediate redshift, several works have shown that the average size of the quiescent ETGs at $0.4 \lesssim z \lesssim 1.5$ is the same for galaxies in the field and dense environments \citeg{Rettura10,Huertas13a,Kelkar15, Saracco17}, and for stellar mass $\mathrm{log}(M/M_{\astrosun}) \sim 10.5-11.8$.
Instead, smaller galaxy sizes in clusters are found by \citealp{Raichoor12} for bulge-dominated galaxies in the Lynx superstructure at $z \sim 1.3$, in the galaxy mass range log$(M/M_{\astrosun}) \sim 10-11.5$), and by \citet{Matharu19} at $z \lesssim 1.5$.

At higher redshift, $z \gtrsim 1.5$, most works find larger quiescent ETG sizes in clusters at both high (\citealp{Papovich12, Strazzullo13, Delaye14, Chan18, Noordeh21}, log$(M/M_{\astrosun}) \sim 10.5-11.5$) and low (\citealp{Mei15}, log$(M/M_{\astrosun}) \sim 9.5-10.5$) mass. These results are also stable when using different galaxy mass proxies \citep{Andreon20}. However, \citet{Allen15}, find larger cluster star-forming galaxies and similar cluster quiescent galaxy sizes compared to the field, in the mass range log$(M/M_{\astrosun}) \sim 9-11.5$. It has to be noticed that their results are limited by their sample size, and they point out that they are only sensitive to differences in size of 0.7~kpc or greater. \citet{Strazzullo22} also find similar cluster quiescent galaxy sizes compared to the field, in  massive galaxy clusters from South Pole Telescope Sunyaev-Zeldovich effect survey at $1.4 \lesssim z \lesssim 1.7$ and galaxy stellar masses of log$(M/M_{\odot}) > 10.85$.




There is also some evidence that the MSR flattens out at low masses, log$(M/M_{\odot})\lesssim 10.3$, in the local Universe \cite{Bernardi11b}, and up to $z\sim1$ \citep{Saracco17, Nedkova21}, as predicted by models \citep{Shankar14a}. 

Finally, several studies find high percentages of compact post-starburst \citep{Maltby18,Socolovsky19,Matharu20,Wilkinson21} and massive compact galaxies \citep{Lu19, Gu20, Tadaki20} in dense environments at $z=1.5-2$.


In this paper, we extended MSR studies in a unique sample of galaxy clusters at redshift $1.4 \lesssim z\lesssim 2.8 $ from the Clusters Around Radio-Loud AGN (CARLA; \citealp{Wylezalek13,Wylezalek14}) survey. We find that passive ETGs in clusters are $\gtrsim 0.2-0.3 {\rm dex}$ larger ($\gtrsim 3\sigma$) than in the field at these redshifts, while late-type galaxies (LTGs) have similar sizes. Combining our results with other cluster studies, we demonstrate that cluster passive ETGs have much slower size evolution than their field counterparts. The brightest bluster galaxies (BCG) lie on the same MSR as satellites. Half of the ETGs with active star-formation lie on the LTGs MSR. The MSR flattens at low mass, and we do not observe large percentages of very compact galaxies in our sample.

%

The structure of this paper is as follows. We describe our observations in Section 2. The galaxy property measurements and our sample selection are presented in Section 3. Our results are presented in Section 4, and discussed in Section 5. Section 6 summarizes the paper. 

Throughout this paper, we adopt a $\Lambda \ $ - cold dark matter ($\rm{\Lambda CDM}$) cosmology with $\Omega_{\rm M} = 0.3$, $\Omega_{\rm \Lambda} = 0.7$, $\Omega_{\rm k} = 0$, and $h = 0.7$, and assume a Chabrier initial mass function (IMF) (\citealp{2003PASP..115..763C}). The photometry and structural parameters in this paper were measured adopting the 3D-HST empirical PSF model \footnote{\url{https://archive.stsci.edu/prepds/3d-hst/}} for the {\it HST}/WFC3 GOODS-S images in the F140W ($H_{140}$) band. Hereafter, we call star-forming galaxies "active", not to confuse with active galactic nuclei.

\section{The CARLA survey} \label{sec:obs}

\subsection{CARLA cluster candidates}

CARLA is a 408h Warm {\it Spitzer} IRAC survey of galaxy overdensities around 420 radio-loud AGN (RLAGN). The AGN were selected across the full sky and in the redshift range of $1.3 < z <3.2$. Approximately half of them are radio-loud quasars (RLQs) and the other half are high-redshift radio galaxies (HzRGs). With the aim to detect galaxy cluster candidates, \cite{Wylezalek13} selected galaxies at $z>1.3$ around the AGN, using a color selection in the IRAC channel~1 ($\lambda =3.6\ \mu {\rm m}$; IRAC1, hereafter) and channel~2 ($\lambda = 4.5\ \mu {\rm m}$; IRAC2, hereafter). They found that $92\%$ of the selected RLAGN reside in dense environments with respect to a field sample in the {\it Spitzer} UKIDSS Ultra Deep Survey (SpUDS, \citealp{Rieke04}), with the majority ($55\%$) of them being overdense at a $> 2\sigma$ level, and $10\%$ of them at a $> 5\sigma$ level. 

From their IRAC luminosity function, \citet{Wylezalek14} showed that CARLA overdensity galaxies have probably quenched faster and earlier than field galaxies. Some of the CARLA northern overdensities were also observed in either deep {\it z}-band or deep {\it i}-band, with Gemini/GMOS, VLT/ISAAC and WHT/ACAM (P.I. Hatch (see below); \citealp{Cooke15}). These observations permitted them to estimate galaxy star formation rate histories, and to deduce that, on average, the star formation of galaxies in these targets had been rapidly quenched (\citealp{Cooke15}).

The twenty highest overdensitiy CARLA {\it Spitzer} candidates were followed by a {\it Hubble Space Telescope} Wide Field Camera 3 ({\it HST}/WFC3) observations (P.I. Stern (see below); \citealp{Noirot16,Noirot18}), and sixteen of them were spectroscopically confirmed at $1.4 < z < 2.8$, together with seven spectroscopically confirmed serendipitous structures at $0.9 < z < 2.1$ \citep{Noirot18}. The structure members were confirmed as line-emitters in H${\alpha}$, H${\beta}$, [{O}{II}], and/or [{O}{III}], depending on the redshift, and have star formation estimates from the line fluxes \citep{Noirot18}. The galaxy star-formation (for stellar mass $\gtrsim 10^{10} M_{\astrosun}$) is below the star-forming main sequence (MS) of field galaxies at a similar redshift. Star-forming galaxies are mostly found within the central regions \citep{Noirot18}. 

Mei et al. (2022, hereafter, M22) performed an in-depth study of the morphology, quiescence and merger incidence of CARLA clusters. They found that the galaxy morphology-density and passive-density relations are already in place at $z\sim 2$. The cluster ETG and passive fractions depend on local environment and mildly on galaxy mass. Active ETGs are 30$\%$ of the total ETG population. Cluster merger fractions are significantly higher than in the CANDELS fields, as predicted from previous studies to explain high quiescent fractions at $z \lesssim 1.5$. Their findings confirm that all the spectroscopically confirmed CARLA overdensities have properties consistent with clusters and proto-clusters.

We describe our observations below.
More details on the {\it Spitzer} IRAC, {\it HST}/WFC3 and ground-based data reduction and results can be found in \cite{Wylezalek13,Wylezalek14}, \cite{Noirot16, Noirot18}, and \citet{Cooke15}, respectively.

\subsection{Spitzer observations}

All CARLA clusters were observed with {\it Spitzer} IRAC1 and IRAC2 (Cycle 7 and 8 snapshot program; P.I.: D. Stern), with total exposure times of 800 s/1000 s in IRAC1 and 2000 s/2100 s in IRAC2, for radio galaxies at $z < 2 / z > 2$, which provided a similar depth in both channels. The IRAC cameras have $256 \times 256$ InSb detector arrays with a pixel size of 1.22 arcsec and a field of view of $5.2 \times 5.2$ arcmin. \citet{Wylezalek13} performed the data calibration and mosaicing with the \texttt{MOPEX} package (\citealp{Makovoz05}) and detected sources with SExtractor (\citealp{sextractor}), using the IRAC-optimized SExtractor parameters from \cite{Lacy05}. The final {\it Spitzer} IRAC1 and IRAC2 mosaic has a pixel size of 0.61~\rm{arcsec}, after taking into account dithering and subpixelation. The 95\% completeness limit is IRAC1=22.6~mag and IRAC2=22.9~mag.

\subsection{HST observations}

The {\it HST}/WFC3 imaging and grism spectroscopy were obtained with a dedicated {\it HST} follow-up program (Program ID: 13740; P.I.: D. Stern). The program consisted of F140W band (hereafter $H_{140}$) imaging with a field of view of $2 \times 2.3 \ \rm{arcmin^2}$ at a resolution of $0.06 \ \rm{arsec \ pix^{-1}}$, after taking into account dithering, and G141 grism spectroscopy with a thoughtput > 10\% in the wavelength range of $1.08 \ \mu {\rm m }< \lambda < 1.70 \ \mu {\rm m }$ and spectral resolution $R = \lambda/\Delta \lambda = 130$. This grism was chosen in order to permit the identification of strong emission lines at our target redshift, such as H$\alpha$, H$\beta$, [{O}{II}] and [{O}{III}]. \citet{Noirot16, Noirot18} performed the data reduction using the \texttt{aXe} (\citealp{Kummel09}) pipeline, by combining the individual exposures, and removing cosmic ray and sky signal. They performed the source detection with SExtractor (\citealp{sextractor}), and extracted two-dimensional spectra for each field, based on the positions and sizes of the sources. The redshifts and emission line fluxes were determined using the python version of \texttt{mpfit} and are published in \cite{Noirot18}. Our {\it HST} image 5$\sigma$ magnitude limit within an aperture of radius of $0.17$~arcsec is $H_{140}=27.1$~mag.

\subsection{Ground-based optical observations}

Ground-based optical imaging in $i$- or $z$-band is available for nine of the CARLA clusters \citep{Cooke15}. Seven clusters were observed in September 2013 - December 2014 using ACAM at 4.2m William Hershel telescope (P.I. Hatch). ACAM has a circular field of view, 8.3 arcmin in diameter with a pixel scale 0.25 arcsec~pixel$^{-1}$. Two other clusters were observed between February and April 2014 with GMOS-S (at the Gemini South telescope) using the EEV detectors. The field of view of GMOS-S is $5.5\times5.5$~arcmin with a pixel scale of 0.146 arcsec~pixel$^{-1}$. Exposure times were calculated depending on the actual seeing, in order to reach a consistent depth across all fields. The reduction of the {\it i}-band images was performed with the publicly available THELI software \citep{Erben05,Schirmer13}. For the photometric calibration we used either available Sloan Digital Sky Survey photometry or standard stars observed before and after the cluster observations. More details on these observations and image reduction can be found in \citet{Cooke15}.
CARLA J2039-2514 has archival imaging observations with VLT/ISAAC (run ID 69.A-0234) in the {\it z}-band with 4800s exposure time (see also \citealp{Noirot16}). 

\section{Sample selection and galaxy property measurements} \label{sec:galprop}

We focus this study on 15 of the 16 CARLA confirmed clusters in \citet{Noirot18}, those that present sufficiently high overdensities to yield low field galaxy contamination (M22). 

\subsection{Galaxy sample selection}
Details on our cluster and galaxy selection are found in M22, and we describe below the main steps leading to our cluster and galaxy sample selection, and the galaxy property measurements.

\subsubsection{Galaxy photometry and mass measurement}

Our photometry was obtained from a joint analysis of IRAC1, IRAC2, $H_{140}$ and, when available, ground-based {\it i}-band or {\it z}-band images. For an efficient source deblending, M22 used the T-PHOT software \citep{merlin+15,merlin+16}, with the high-resolution {\it HST} images as priors to derive PSF-matched fluxes in the lower-resolution bands. 

M22 measured our CARLA galaxy stellar masses by calibrating our PSF-matched {\it Spitzer}/IRAC1 magnitudes with galaxy stellar masses from \cite{Santini15} derived from the \cite{guo13} multiwavelength catalog in the CANDELS WIDE GOODS-S field. Hereafter, we use the symbol $M$ for the galaxy stellar mass.
The {\it Spitzer} IRAC1 magnitudes correspond to the rest-frame near-infrared in the redshift range of the CARLA sample, and they expected them not to be biased by extinction. M22 found a very good correlation between these magnitudes and the \cite{Santini15} mass measurements, with scatters of $\approx 0.12$~dex at the redshift of the clusters studied in this paper. Adding in quadrature the scatter of the relation and uncertainties from \cite{Santini15}, they obtained mass uncertainties in the range $\sim 0.4-0.5$~dex, and $\approx 0.2-0.3$~dex for $9.6 \lesssim $ log$_{10} (M/M_{\astrosun}) <10.5$ and log$_{10} (M/M_{\astrosun}) > 10.5$, respectively. 

\begin{table}
\caption{SExtractor parameters used for source detection.}
\label{tab:setup_se}
\begin{tabular}{lcc}
\hline\hline
SExtractor & Cold Mode & Hot Mode\\
\hline
DETECT\_MINAREA & 5.0 & 10.0 \\
DETECT\_THRESH & 0.75 & 0.7 \\ 
ANALYSIS\_THRESH & 5.0 & 0.8 \\
FILTER\_NAME & tophat\_9.0\_9x9 & gauss\_4.0\_7x7 \\
DEBLEND\_NTHRESH & 16 & 64 \\
DEBLEND\_MINCONT & 0.0001 & 0.001 \\
BACK\_SIZE & 256 & 128 \\
BACK\_FILTERSIZE & 9 & 5 \\ 
BACKPHOTO\_THICK & 100 & 48 \\
\hline
\end{tabular}
\end{table}

\begin{table}
\caption{Constraints on GALFIT parameters.The constraint on magnitude is relative to the measured SExtractor magnitude.}
\label{tab:setup_ga}
\begin{tabular}{cccc}
\hline\hline
parameter & description & constraints & units \\
\hline
$n$ & S\'ersic index & 0.2 : 8 & \\
$R_e$ & effective radius & 0.3 : 400 & pixel \\
$Q$ & axis ratio & 0.0001 : 1 &\\
$m$ & magnitude & -3 : +3 & SExtractor mag\\
\hline
\end{tabular}
\end{table}

\subsubsection{Sample selection}

M22's sample selection aims at optimizing completeness and purity. 
Observations of most of the CARLA clusters and proto-clusters include three ($H_{140}$, IRAC1, IRAC2) to five bandpasses (ground based {\it i}-band and {\it z}-band, $H_{140}$, IRAC1, IRAC2), and they could not perform a precise photometric redshift analysis from their spectral energy distribution. Instead, they selected galaxies in color and spatial regions where they expected a low outlier contamination. 

M22 selected galaxies with $(\mathrm{IRAC1} - \mathrm{IRAC2})>-0.1$, $\mathrm{IRAC1} <22.6$~mag, from which they obtained a sample $\sim90\%$ pure and complete for galaxies at $z>1.3$. To reduce the contamination from outliers with $z>1.3$, but not at the cluster redshift, they only selected galaxies located in the densest cluster regions, in circles of radius of 0.5~arcmin ($\sim 0.25$~Mpc at our redshifts), in which the background contamination is $\lesssim 20\%$ in most clusters (M22). The scale of these regions corresponds to the scale of the dense cluster cores at $z \sim 1$ \citep{Postman05}. 

They also select galaxies brighter than $H_{140}=24.5$~mag. In fact,
\citet{vanderWel12} and \citet{Kartaltepe15} showed that morphological classification and the measurement of galaxy structural parameters are dependable only for magnitudes brighter than the WFC3/F160W magnitude $H_{160}=24.5$~mag in the CANDELS Wide survey \citep{koekemoer11}. The CANDELS Wide survey reaches a 5$\sigma$ magnitude limit of $H_{160}=27.4$~mag, which is comparable to the CARLA magnitude limit of $H_{140}=27.1$~mag (both were calculated within an aperture with $0.17$~arcsec radius), when the different filter response functions are taken into account. For this reason, in this paper, we did not perform further simulations to assess the precision and bias of our measurements, and rely on the finding from \citet{vanderWel12} for the choice of the magnitude limit of the galaxy sample chosen for our work. \citet{vanderWel14} also pointed out that structural parameters measurements performed in infrared band-passes at our redshifts do not show significant differences.

M22's final sample includes a total of 271 galaxies in fifteen CARLA confirmed clusters and nineteen overdense regions. In fact, three of our clusters are double structures (CARLAJ1358+5752, CARLAJ1018+0530, and CARLAJ2039-2514), as predicted by cosmological models for clusters assembling at $z=1.5-3$ \citep{Chiang13, Muldrew15}.
Galaxies that were spectroscopically confirmed at a redshift different than the clusters by \citet{Noirot18} and a recent photo-spectral analysis of CARLAJ1018+0530 by Werner et al. (MNRAS, submitted) were not included in the final catalog.

\subsubsection{Galaxy morphological classification and passive and active galaxy selection}


M22 performed a galaxy visual morphological classification using two large morphological classes, ETGs and late-type galaxies (LTGs). ETGs include spheroid and compact galaxies, and LTGs include disks and irregular galaxies. These correspond to the main morphological classes used in the CANDELS survey \citep{Kartaltepe15}:
(1) disk, these galaxies have a disk even if they do not show clear spiral arms;
(2) spheroid, these galaxies are resolved spheroids and do not show a disk;
(3) irregular, all extended galaxies that can be classified neither as a disk nor as a spheroid;
(4) compact and unresolved, these are compact or unresolved galaxies;
(5) unclassifiable.
The sample used in this paper does not include any unclassifiable galaxies. 


Nine of our CARLA clusters have been observed in the {\it i} or {\it z}-band from the ground (M22), which correspond to a rest-frame U/NUV band. For these clusters, M22 identified passive and active galaxies using color-color diagrams, which correspond to the {\it UVJ} diagrams used in the literature to separate passive from active dusty galaxies up to a redshift $z=3.5$ \citeg{Labbe05,Wuyts07,Williams09,Whitaker11,Fang18}. The fraction of the galaxies that could not be classified as passive and active (because of the lack of ground-based observations) corresponds to $\sim 30$\% of the galaxies. They selected passive galaxies as galaxies with specific star formation rate $\mathrm{log(sSFR)}<-9.5$~yr$^{-1}$, using the CANDELS \citet{Santini15} sSFR as the reference for their selection calibration. This selection permitted them to obtain passive samples that are $\sim$80-85\% complete and pure, and includes recently quenched galaxies at $\sim3 \sigma$ below the field star formation main sequence. 




\begin{figure}
	\centering
	\includegraphics[width=1\hsize]{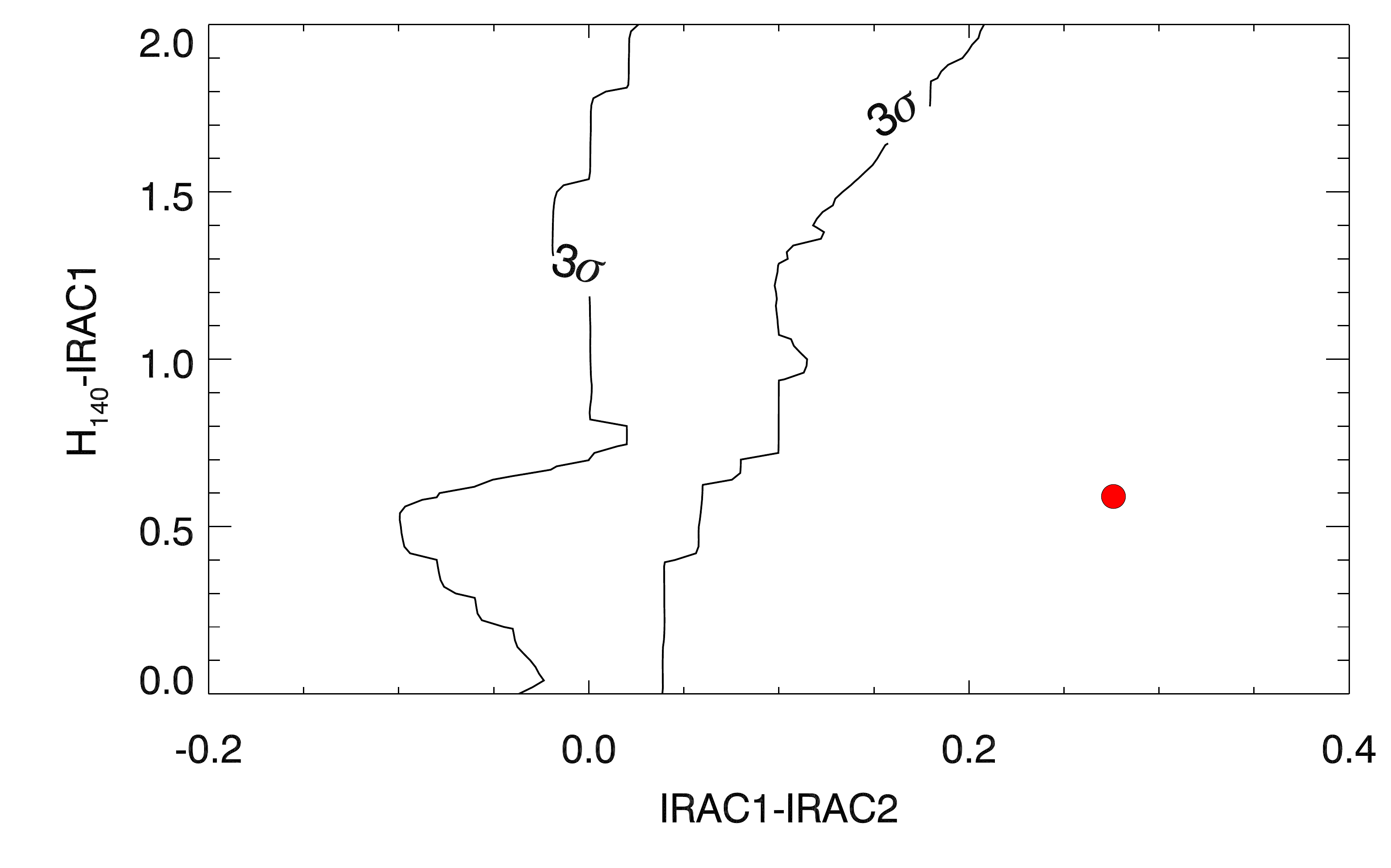}
		\caption{Two color diagram to separate stars and galaxies. The $3\sigma$ external locus of the modeled star distribution is shown by a black contour, the unresolved object from our sample is shown in red, and is most probably extragalactic.}
		\label{pic_all}
\end{figure}
 
\begin{figure*}[ht!]
\center
 \includegraphics[width=0.289\textwidth]{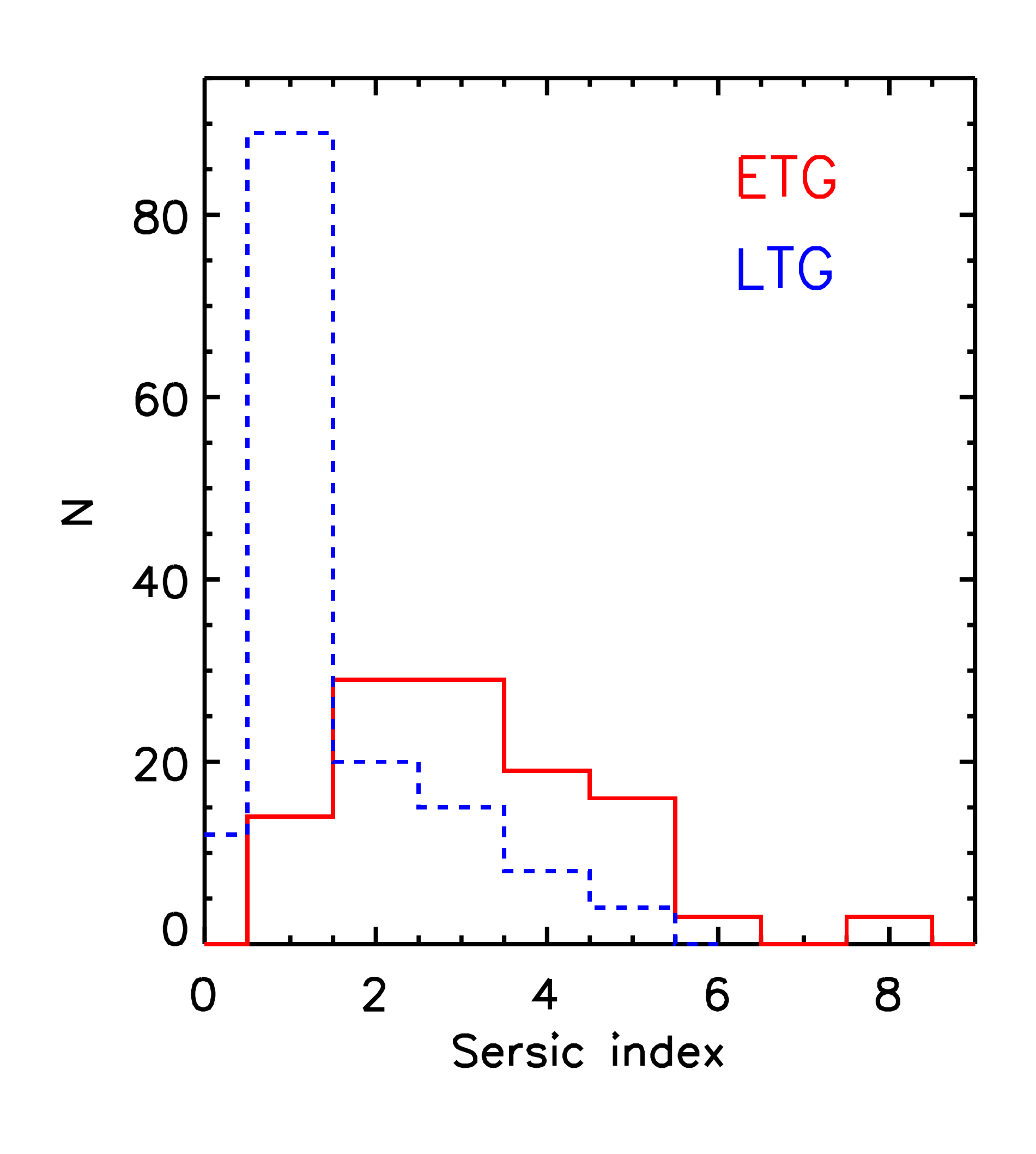}
\includegraphics[width=0.285\textwidth]{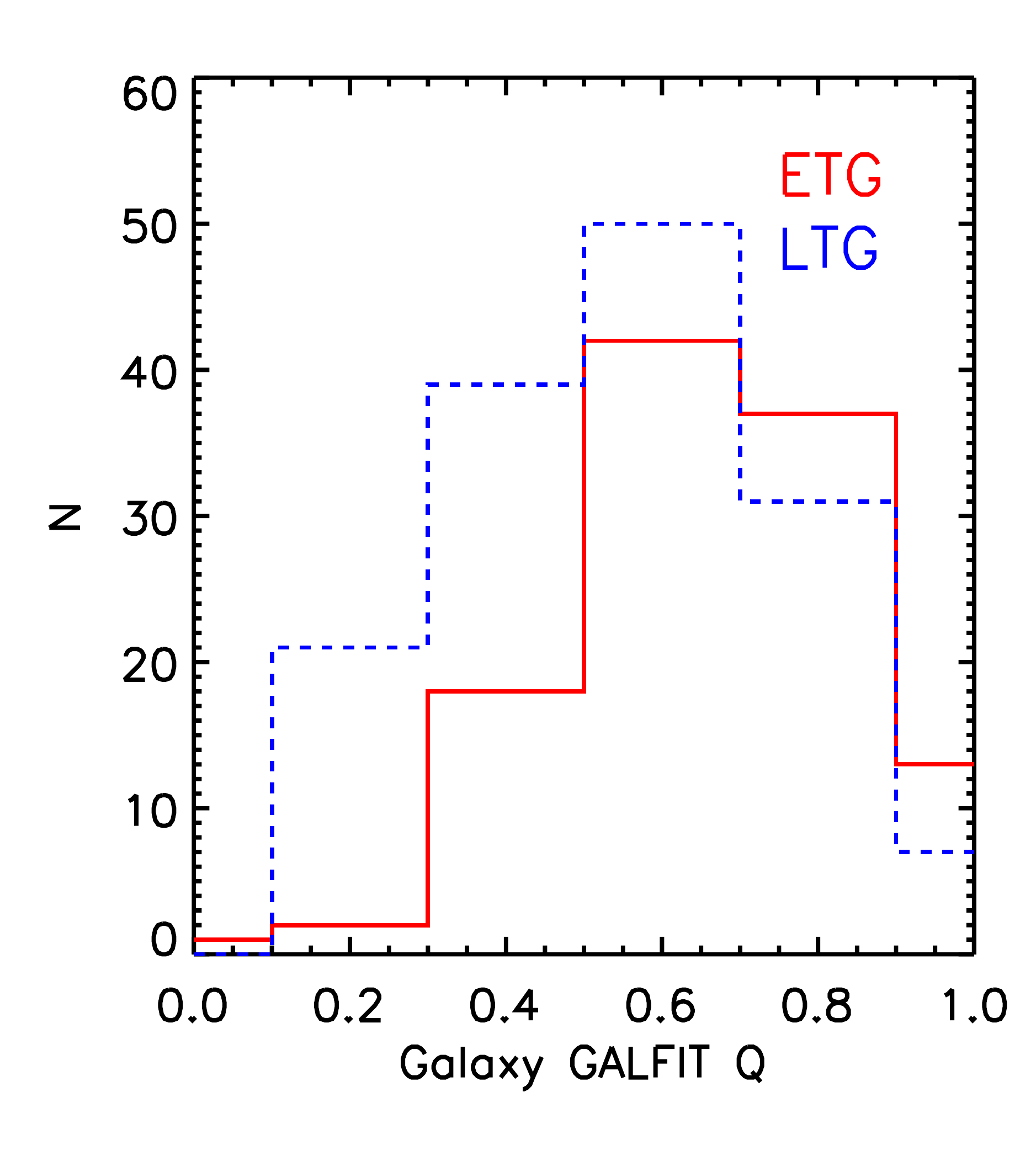}
\includegraphics[width=0.287\textwidth]{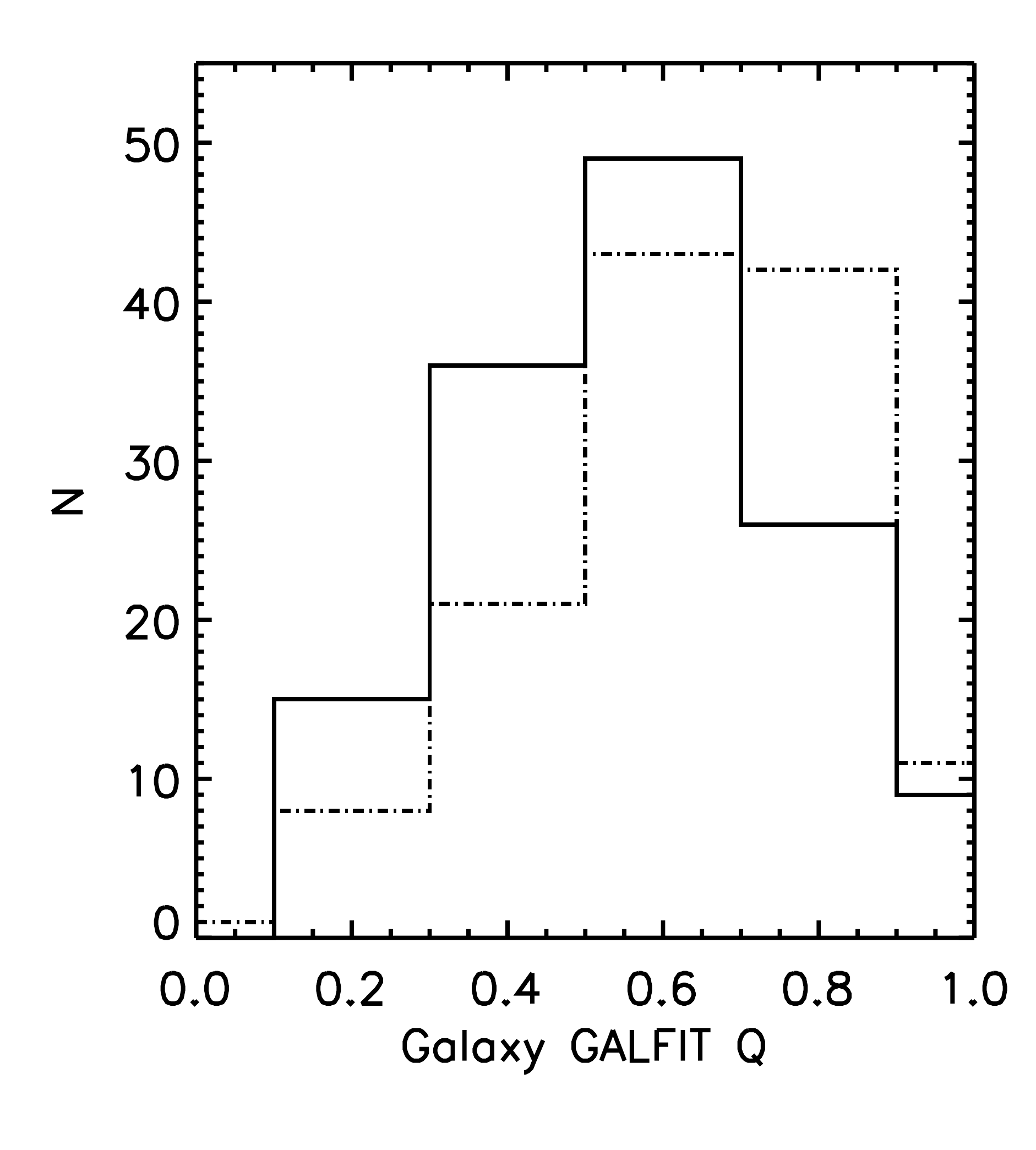}
\caption{CARLA cluster structural parameter distribution for our main morphological classes: ETG (red solid line) and LTG (blue dashed line). Left: S\'ersic index distribution; Middle: GALFIT $Q$ distribution; Right: GALFIT Q distribution for asymmetric (solid line) and symmetric galaxies (dash-dotted line). Our parameters are consistent with the visual morphological classification. }\label{fig:strpar}
\end{figure*} 

\subsection{Measurements of galaxy structural properties} 


We measured galaxy structural parameters using the software GALAPAGOS \citep{barden12}, using the high-resolution $H_{140}$ images. The $H_{140}$ channel corresponds to the rest-frame V band in all but the highest redshift cluster in our sample, CARLA J1017+6116, where $H_{140}$ instead corresponds to the rest-frame U-band. GALAPAGOS performs the following main steps:
source detection;
creation of image and noise cutouts for each detected source;
estimation of the local background;
fit of the surface brightness profile to a S\'ersic profile; 
and compilation of all objects into a final output catalog.

The source detection is based on SExtractor \citep{sextractor}. Following \citet{barden12}, we ran SExtractor on the $H_{140}$ images in the {\it cold} and {\it hot} modes, which are optimized to detect bright and faint objects, respectively. We adopted the same configuration of parameters used for the catalogs released by \citep[CANDELS,][]{grogin11,koekemoer11}, and published by \citet{galametz13} and \citet{guo13}. More specifically, we created a first catalog including all the {\it cold} sources; then we compared every source detected in the {\it hot} mode to the first catalog detections, added those whose central position did not lie inside the Kron ellipse of any {\it cold} source, discarding the others. Table \ref{tab:setup_se} shows the key SExtractor parameters used in our source detection.

The photometric and structural parameter estimation was based on GALFIT \citep{peng02}, which fits the surface brightness profile of each detected source
to a one-component {\em S\'ersic} model \citep{sersic}, 
defined by the following free parameters:
the total magnitude $m$, the half-light radius measured along the major axis (effective radius) $R_e$, the S\'ersic index $n$, the axis ratio $Q$ (the ratio between the model minor and major axis, $b/a$), the position angle $P.A.$, and the central position. The software uses the Levenberg-Marquardt algorithm to minimise the residual between a galaxy image and the PSF-convolved model by modifying the free parameters. 

We used the same GALFIT configuration as \citet{vanderWel12, vanderWel14} (Table \ref{tab:setup_ga}), given that the CANDELS Wide survey depth is comparable to the CARLA depth (see Sec.~3.2), and to have homogeneous measurements of cluster and field sizes.
The conversion to physical length-scale in kiloparsec was performed using the angular distance of each cluster, assuming that all cluster galaxies have the same redshift as the average redshift of the cluster from \citet{Noirot18}. 


We reran GALFIT on the 22 galaxies for which either GALAPAGOS did not converge or the resulting values had uncertainties greater than the value itself, or the parameters hit the constrains set for either $R_e$, $n$ or $Q$. In those cases, we tried different values of the input parameters to find a stable global minimum of the residuals and resolve the problems listed above. We divided our sample in 3 categories: (i) Galaxies with a good quality fit; (ii) QSO, whose effective radii are uncertain due to saturation in the {\it HST} image (9 objects); (iii) Unresolved galaxies, where GALFIT converged close to the minimum constraint for the effective radius. We exclude QSO from our analysis.

The unresolved galaxy category consists of only one object with an effective radius $R_e=0.5$~pix, which is close to the lower limit for the $R_e$ estimate used by \citet{vanderWel12} ($R_{e,min}=0.018$ arcsec or 0.3 pix in WFPC3 image). The objects with such small effective radii are essentially indistinguishable from point sources, so their $R_e$ is an upper limit, and they might be either a bona-fide extragalactic object, or a Milky Way star. This object is not listed in the {\it Gaia} EDR3 catalog as a star \citep{GaiaEDR3}.
Additionally, we used the TRILEGAL code\footnote{http://stev.oapd.inaf.it/cgi-bin/trilegal} \citep{Girardi05} to obtain a sample of simulated stars with magnitudes IRAC1$ < 26$~mag at the source RA and DEC, and with standard settings for the geometry of the thin disk, the thick disk and the halo of the Milky Way, as well as for their stellar population parameters. 
We built a ($H_{140}$-IRAC1) vs (IRAC1-IRAC2) diagram (see Fig.~\ref{pic_all}), to identify the locus of the synthetic star colors (see Fig.~\ref{pic_all}). The source does not lie in the star locus, and we keep it as a bona-fide galaxy.

In Fig.~\ref{fig:strpar}, we compare our GALAPAGOS structural parameters with visual morphology from M22. The median S\'ersic index for ETGs and LTGs is $\sim$3 and $\sim$1, respectively, consistent with what is expected for ETG de Vaucouleur and LTG exponential profiles. The median $Q$ for early, late, asymmetric and symmetric galaxies is $\sim$~0.7, 0.55, 0.7, and 0.55, respectively, with early and symmetric galaxies being rounder, as expected.

\begin{figure*}[!]
	\centering
	\includegraphics[width=0.75\hsize]{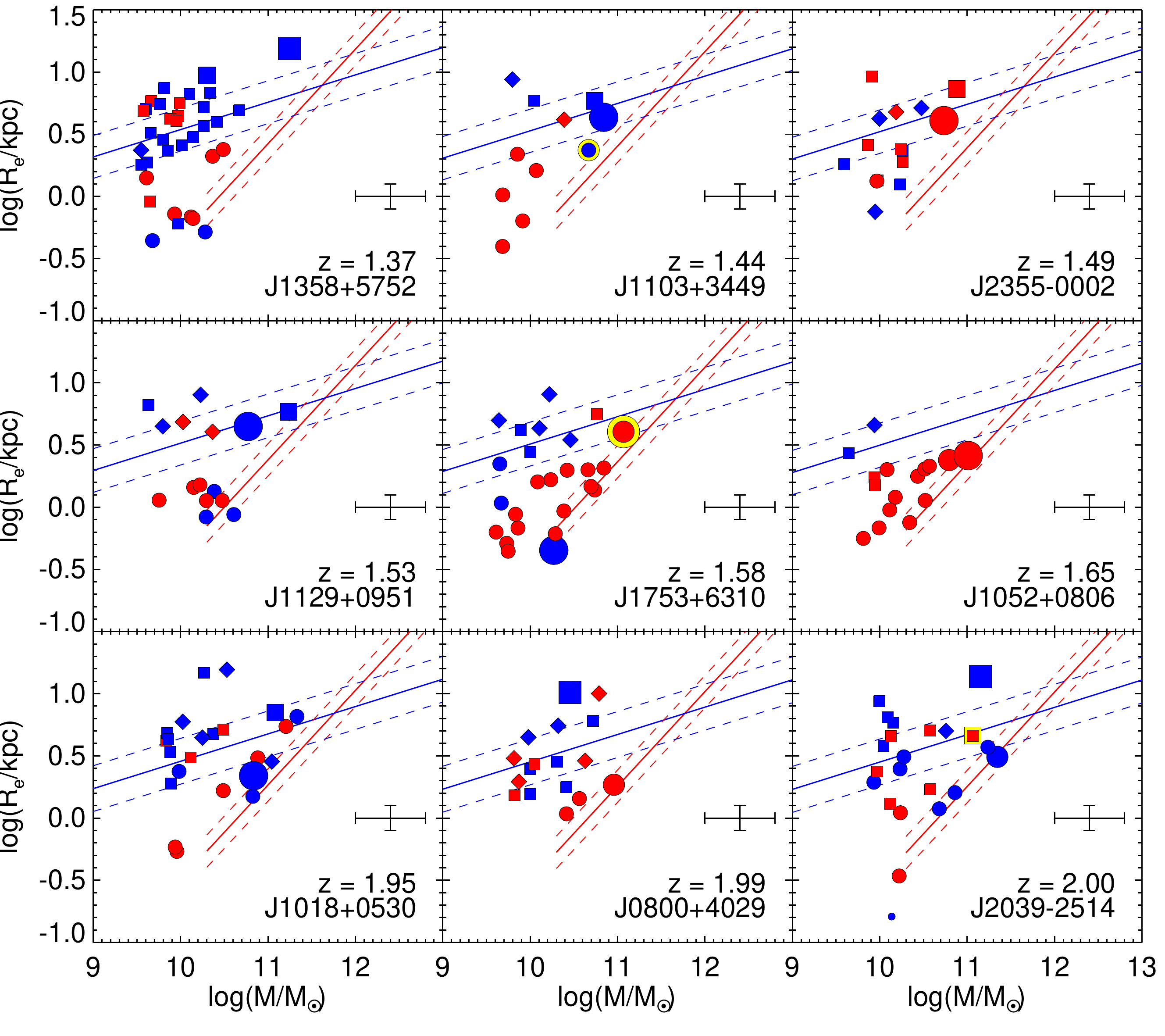}
		\includegraphics[width=0.75\hsize]{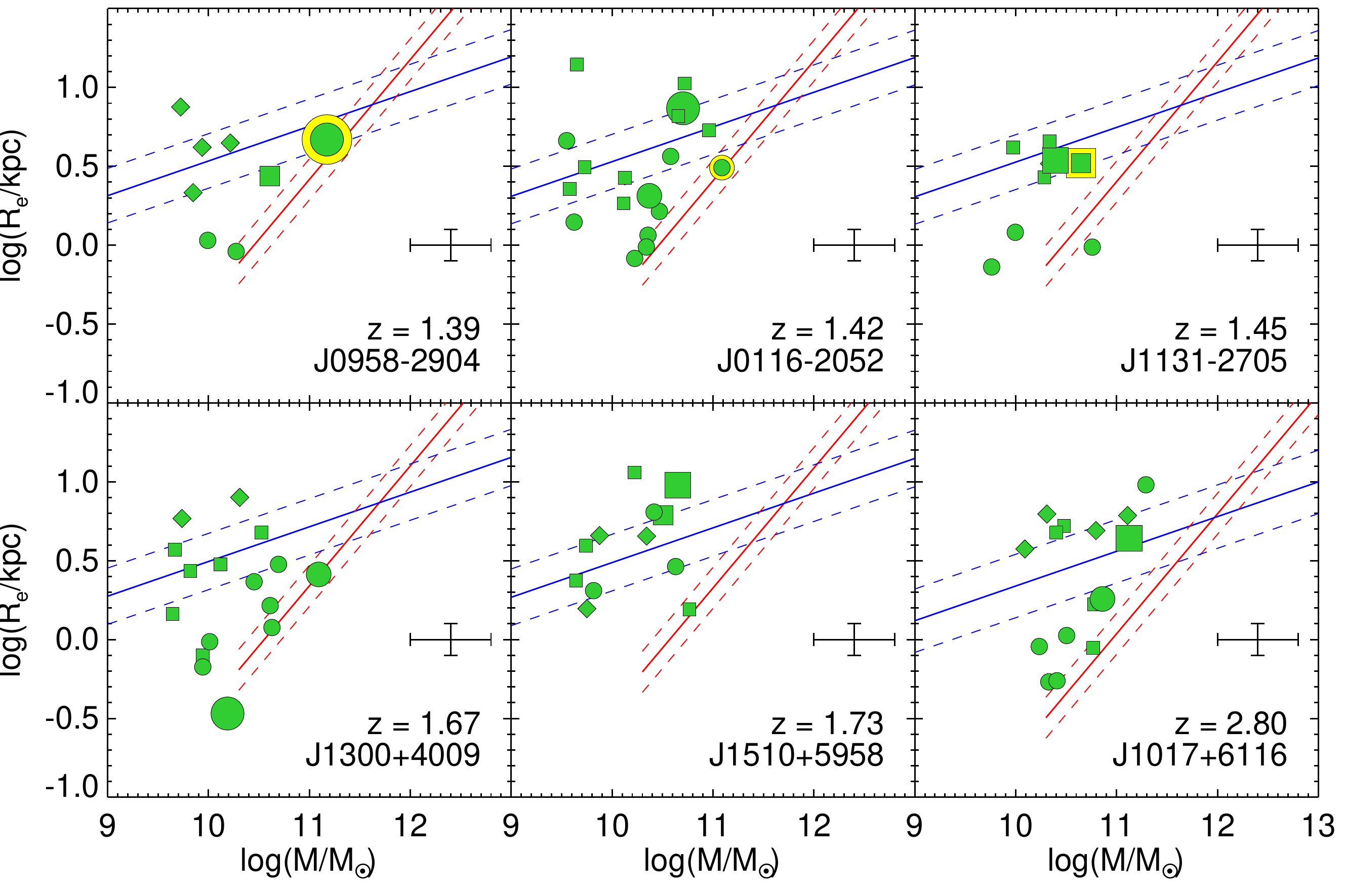}
		\caption{The MSRs for each cluster. Red and blue colors show passive [${\rm log(sSFR)}<-9.5$~yr$^{-1}$] and active galaxies, respectively. Green symbols show galaxies for which we could not separate passive from active galaxies. Circles, squares, and diamonds correspond to ETGs, LTGs and irregulars, respectively. The largest symbols show the BCGs the second largest indicate the second brightest galaxies. The compact galaxy in the J2039-2514 is shown by the smaller symbol. A yellow halo around the galaxy symbol indicates a HzRG. The red and blue solid lines are CANDELS \citet{vanderWel14}'s MSRs for passive and active galaxies, respectively, interpolated to the redshift of each cluster. The CANDELS passive galaxy MSR is shown at log$(M/M_{\odot})>10.3$ to reflect the fitting range in \citet{vanderWel14}. The dashed lines are 1-$\sigma$ scatter for these relations. The average measurement uncertainty is provided on the right-hand side of each box. Cluster active galaxies and LTGs lie on the \citet{vanderWel14}'s active galaxy MSR. Cluster passive ETGs have systematically larger sizes than \citet{vanderWel14}'s passive galaxies.}
		\label{mass-size}
\end{figure*}

\begin{figure*}[!t]
	\centering
	\includegraphics[width=0.45\hsize]{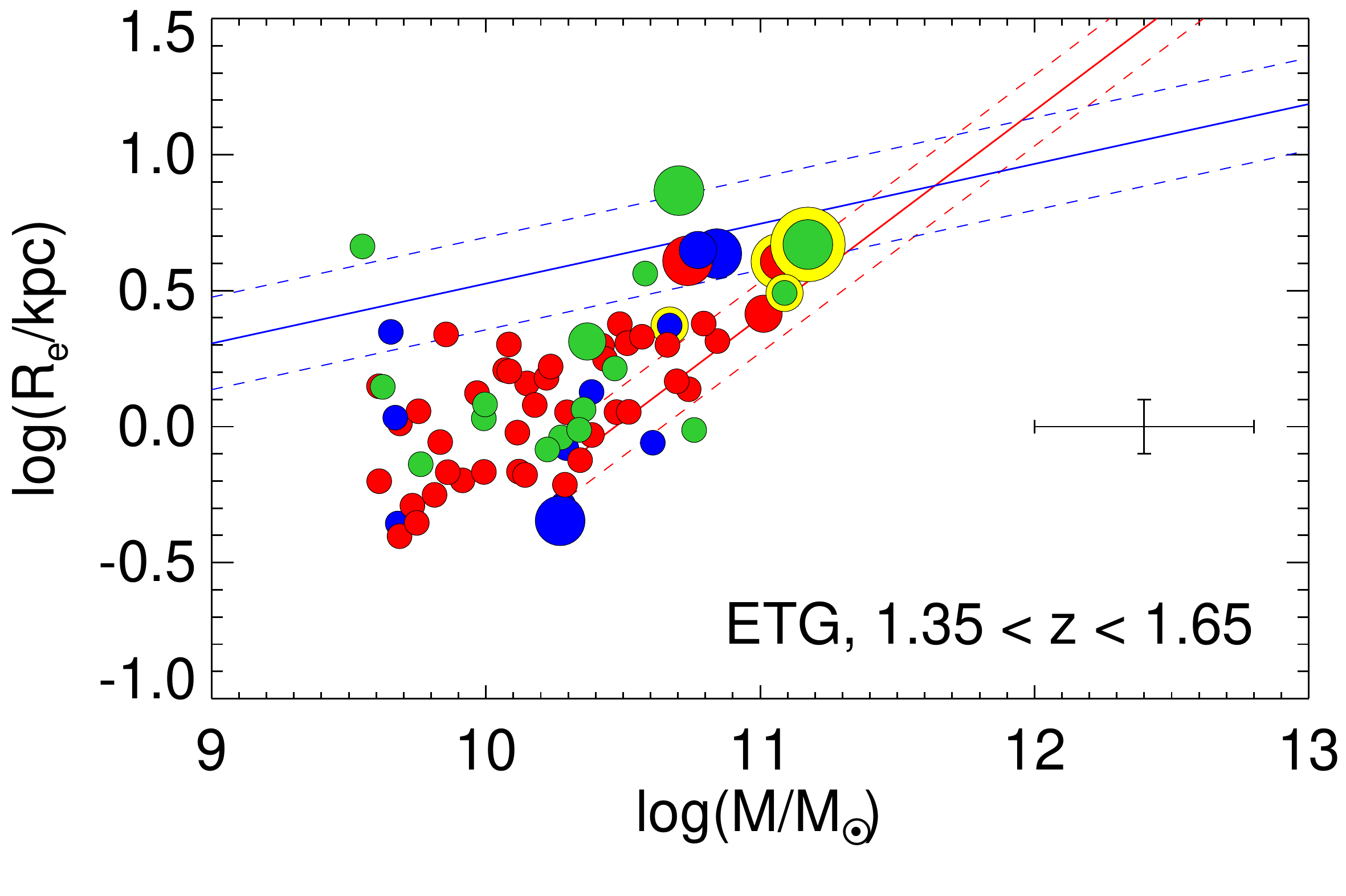}
	\includegraphics[width=0.45\hsize]{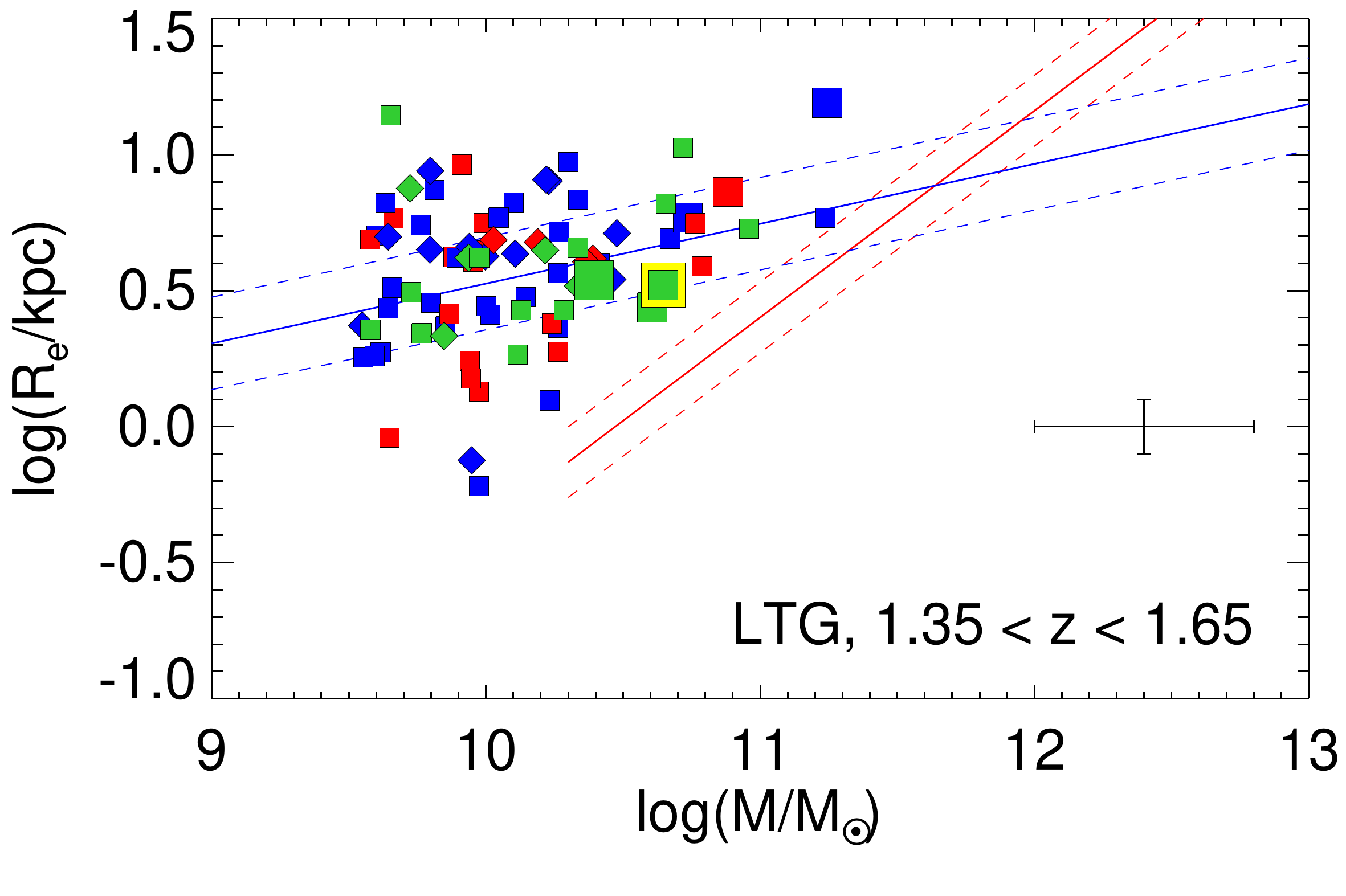}
	\includegraphics[width=0.45\hsize]{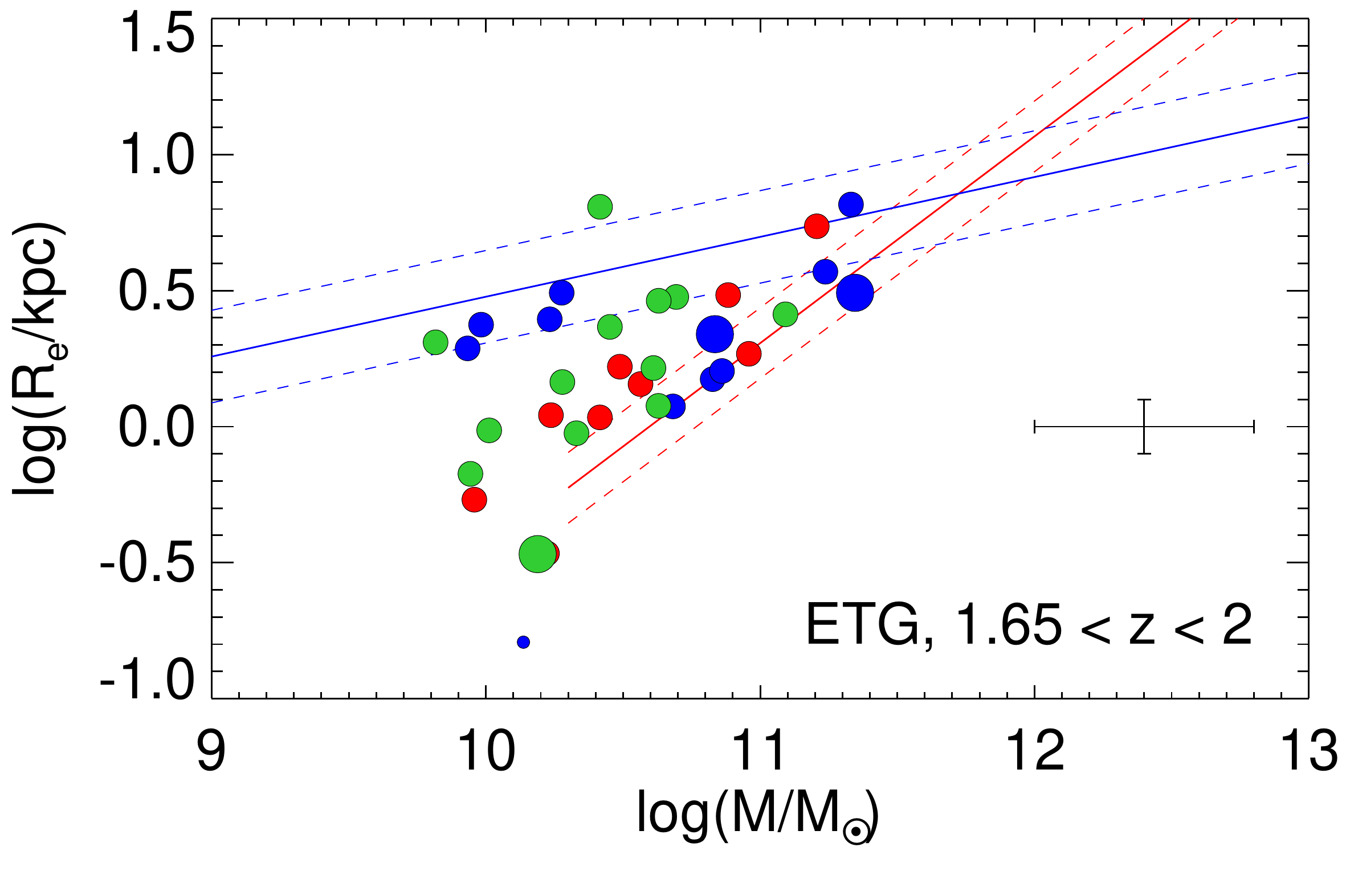}
	\includegraphics[width=0.45\hsize]{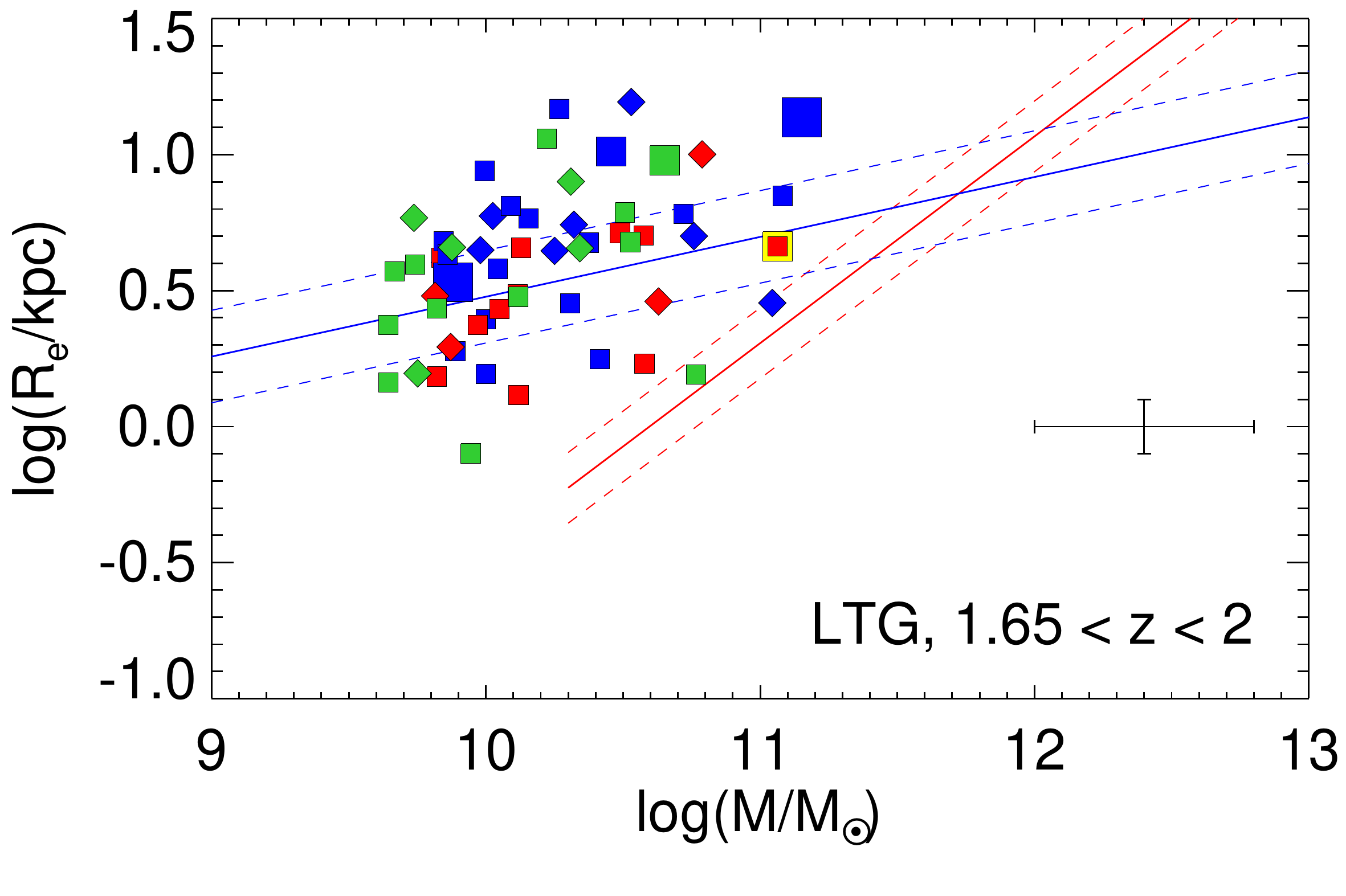}
	\caption{The MSR in two redshift bins. The symbols are the same as in Figure \ref{mass-size}. The \citet{vanderWel14}'s MSRs are shown at the mean redshift of each bin ($z=1.47$ and $z=1.80$). The CANDELS passive galaxy MSR is shown at log$(M/M_{\odot})>10.3$ to reflect the fitting range in \citet{vanderWel14}. The average measurement uncertainty is provided on the right-hand side of each box. In both redshifts bins, cluster active and LTG galaxies lie on the \citet{vanderWel14} active galaxy MSR. Cluster passive ETGs have systematically larger sizes than \citet{vanderWel14}'s passive galaxies. }
		\label{pic_cumul_nosfr}
\end{figure*}

\section{Results}\label{sec:res}
Fig.~\ref{mass-size} shows each cluster passive and active galaxy MSR, compared to CANDELS \citep{vanderWel14}. We interpolated the \citet{vanderWel14} relations at each cluster redshift. While the active and LTG distributions lie on the same active galaxy MSR as \citet{vanderWel14}, the passive and ETG population systematically lie above the \citet{vanderWel14}'s passive galaxy relation. This is also shown in Fig.~\ref{pic_cumul_nosfr}, where we divide the sample in two redshift bins and observe a similar behavior. The relation also indicates a tendency to flatten at $\mathrm{log}(M/M_{\odot})\lesssim 10.5$. 

About $\sim 30\%$ of the cluster ETG are active, and mostly lie on the LTG galaxy MSR. The bulk of these active ETGs is found in just two clusters (J1018 and J2039, both around $z\sim2$; M22). 

The cluster BCG and the second brightest are shown with larger symbols, and lie on the same MSR as the satellites.

\begin{figure*}
	\centering
	\includegraphics[width=0.8\hsize]{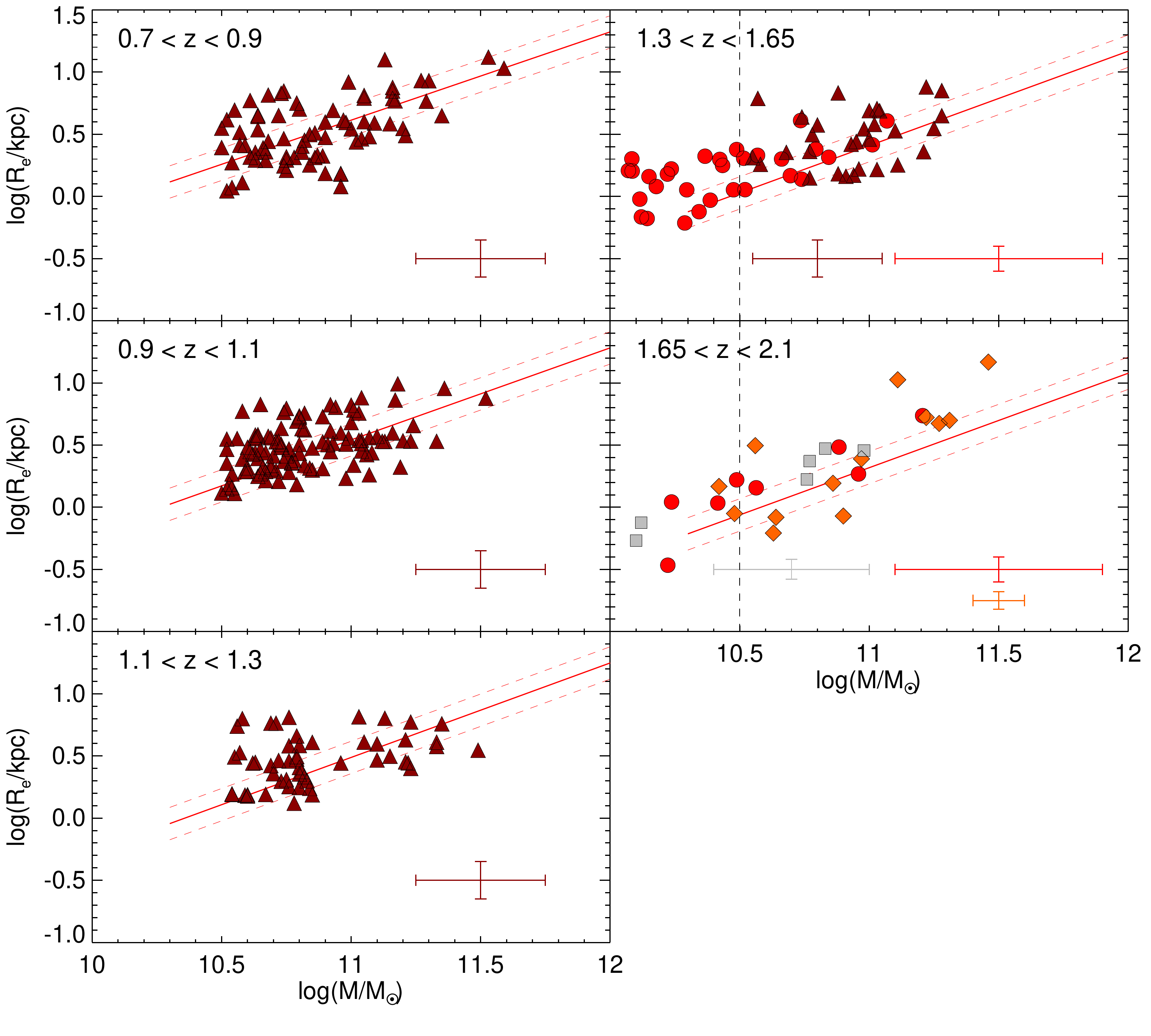}
		\caption{Cluster passive ETG MSRs compared to the field MSRs \citep{vanderWel14} in different redshift bins. The red circles are the CARLA sample (this paper). The brown triangles, gray squares, and orange diamonds are observations from \citet{Delaye14}, \citet{Strazzullo13}, and \citet{Newman14}, respectively. The red continuous line is the \citet{vanderWel14} CANDELS passive galaxy relation, and the dashed lines show the 1~$\sigma$ uncertainty. The vertical black dashed line shows the lower limit of our high mass sample. The average data uncertainties are shown in their corresponding color in the bottom-right corner of each subplot. Cluster passive ETG are systematically larger than CANDELS passive galaxies. }
		\label{pic_evol}
\end{figure*}

\begin{figure*}
	\centering
	\includegraphics[width=0.8\hsize]{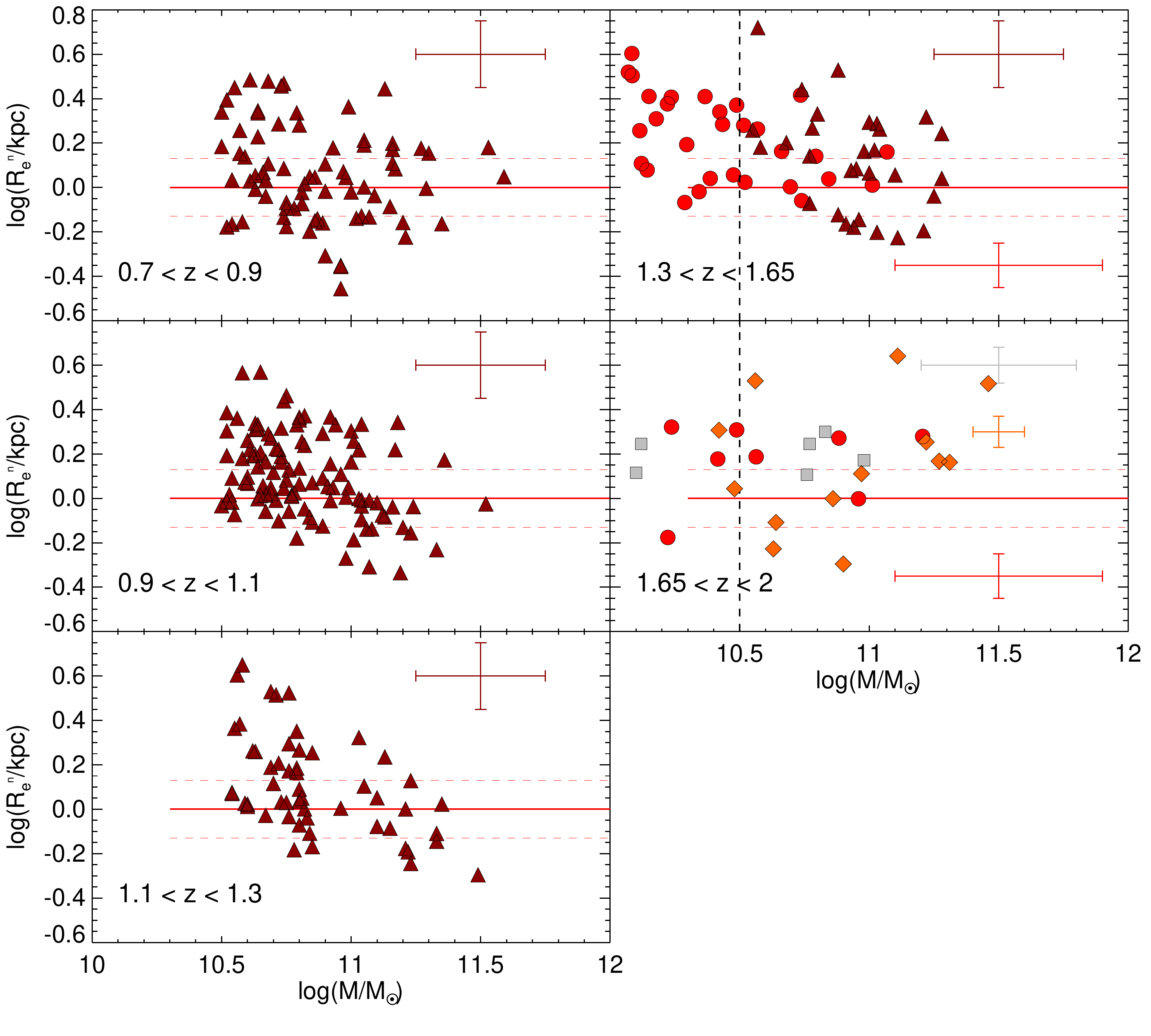}
		\caption{Effective radii normalized to the passive MSR from \citet{vanderWel14}, $R_e^n$, for the same observations as Fig.~\ref{pic_evol}. The symbols are the same as in Figure~\ref{pic_evol}. The average data uncertainties are shown in their corresponding color on the right-hand side of each subplot. }
		\label{pic_evol_norm}
\end{figure*}

Fig.~\ref{pic_evol} shows the evolution of the passive ETG mass-size relation in clusters in the redshift range $0.7<z<2$ compared to the CANDELS \citep{vanderWel14} MSR. We added to our sample the passive ETG observations from \citet{Strazzullo13}, \citet{Delaye14}, and \citet{Newman14}, which used analyses similar to ours. When authors published circularized effective radii, defined as $R_{e,circ}=R_e \sqrt{(b/a)}$ \citeg{Delaye14}, we convert their sizes to the S\'ersic profile half-light radii along the major axis. Fig.~\ref{pic_evol_norm} shows effective radii normalized to the passive MSR from \citet{vanderWel14}, $R_e^n$, for the same observations as Fig.~\ref{pic_evol}. Cluster galaxy sizes are on average larger.

\begin{figure*}
	\centering
	\includegraphics[width=0.7\hsize]{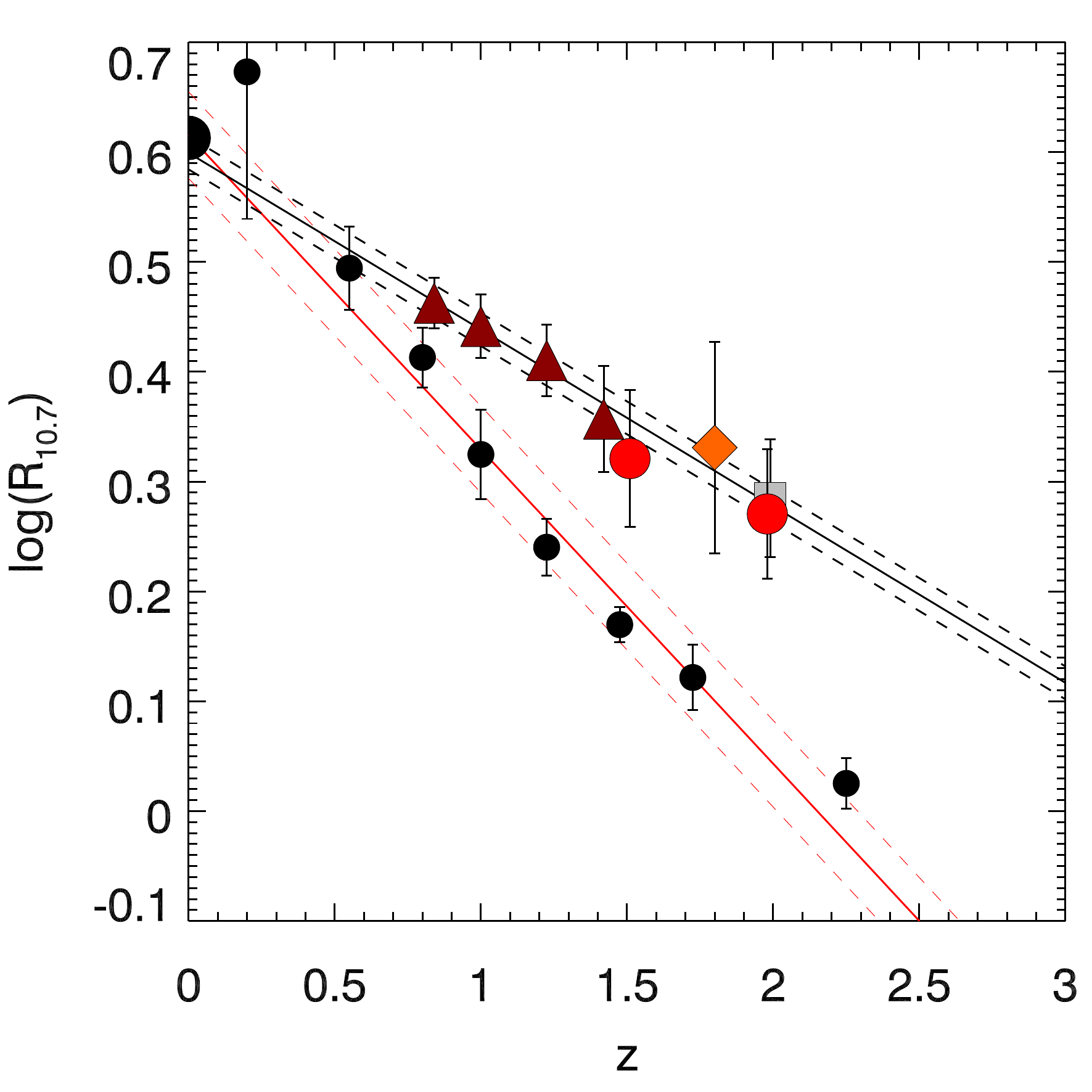}
		\caption{ The evolution of the passive ETG mass-normalized radius $R_{10.7}$ (see text) with redshift. The red circles are the CARLA sample (this paper).	The brown triangles, gray squares, and orange diamonds show observations from \citet{Delaye14}, \citet{Strazzullo13}, and \citet{Newman14}, respectively. The black circles are field ETG sizes taken from \citet{Bernardi14} and \citet{Huertas13b} for $z=0$, and from \citet{vanderWel14} for the other redshifts.
		}
		\label{pic_mei15}
\end{figure*}

To better visualize the difference between cluster and field average sizes and study size evolution, Fig~\ref{pic_mei15} shows the redshift evolution of the mass-normalized radius $R_{10.7}$, defined as \citep{vanderWel14} :
\begin{equation}
R_e~(\mathrm{kpc})=R_{10.7}~(\mathrm{kpc}) \left( \frac{M}{5 \times 10^{10}M_{\astrosun}} \right)^\beta.
\end{equation}
In the conversion, we use the slope $\beta \sim 0.74 - 0.76$ from \citet{vanderWel14}, interpolated to the redshifts that we are considering. Here we compute the average log$(R_{10.7}/\mathrm{kpc})$ in each redshift bin for \citet{Delaye14} and our data, and in each cluster for \citet{Strazzullo13} and \citet{Newman14}. The \citet{Delaye14} observations are averaged over the same redshift bins ($0.7<z<0.9$; $0.9<z<1.1$; $1.1<z<1.3$ and $1.3<z<1.6$) as in Figure~\ref{pic_evol}. Our CARLA cluster observations are averaged over two redshift bins: $1.35<z<1.65$, and $1.9<z<2$. 
The uncertainties on $R_{10.7}$ are calculated using Monte Carlo simulations. 
For this figure, we only consider galaxies with $\mathrm{ log} \left(M/M_{\astrosun} \right) > 10.5$ for a homogeneous sample comparison. We select objects from \citet{vanderWel14} applying the same color and magnitude cuts in IRAC that we applied to our sample.

While the MSR of cluster and field passive ETGs is mostly similar in the local Universe \citeg{Huertas13b}, cluster ETG sizes are systematically larger than field passive galaxies for $\mathrm{ log} \left(M/M_{\astrosun} \right) > 10.5$ and $z>1$ and their evolution is slower in the range $1 \lesssim z \lesssim 2$. In fact, while the cluster and field MSR are superposed within $\sim 1 \sigma$ (Fig.~\ref{pic_evol}), the average normalized cluster radii are $ \gtrsim 3 \sigma$ larger than the field (Fig~\ref{pic_mei15}). We quantify this difference also by fitting the redshift evolution of the cluster $R_{10.7}$ for galaxy mass $\mathrm{ log} \left(M/M_{\astrosun} \right) > 10.5$:
\begin{equation}
{\rm log}(R_{10.7}/\mathrm{kpc})=(-0.16 \pm 0.02) \ (z-1)+(0.44 \pm 0.01),
\end{equation}
compared to the evolution in the field from \citet{vanderWel14}: 
\begin{equation}
{\rm log}(R_{10.7}/\mathrm{kpc})=(-0.28 \pm 0.04)\ (z-1)+(0.33 \pm 0.02).
\end{equation}
The fit was performed by taking into account the uncertainties on both axes, and the uncertainties on the fit are quantified with Monte Carlo simulations.
For galaxies with $\mathrm{ log} \left(M/M_{\astrosun} \right) > 10.5$, at $1<z<2$ cluster passive ETG are on average $>0.2-0.3 {\rm dex}$ ($\gtrsim 3\sigma$) larger than the field. At $z=1.5$ the cluster passive ETGs are $\sim$40\% larger than passive galaxies in the field, and at $z=2$ they are larger by $\sim$120\%. The passive ETG fraction in clusters is $\sim 60\pm10\%$, compared to $\sim 28\pm2\%$ in CANDELS. 

At lower mass, $\mathrm{ log} \left(M/M_{\astrosun} \right) < 10.5$, the MSR is predicted \citep{Shankar13} and observed \citep{graham06,Lange15,li18,Hamadouche22} to flatten at least up to $z=2$ \citep{Nedkova21}. In the range $ 9.6<\mathrm{ log }\left(M/M_{\astrosun} \right) < 10.5$ , we measure an average cluster passive ETG size of $\mathrm{log}(R_e/\mathrm{kpc}) = 0.05 \pm 0.22$.

\section{Discussion}\label{sec:disc}

We observe larger passive ETG sizes in CARLA clusters ($1.4 \leq z \leq 2.8$) when compared to CANDELS field passive ETGs \citep{vanderWel14}. Larger passive ETG sizes in clusters at $z>1$ are also observed by \citet{Strazzullo13}, \citet{Delaye14}, \citet{Newman14}, and \citet{Andreon20} in similar galaxy mass ranges. We obtain the evolution of the mass-normalized radius $R_{10.7}$ as a function of redshift, which shows that passive ETGs with $\mathrm{ log} \left(M/M_{\astrosun} \right) > 10.5$ and $z>1$ are systematically larger in clusters than in field environment. Their mass growth at $ 1\lesssim z \lesssim 2$ is slower than in the field.

\subsection{The mass-size relation at $z\sim 1$}

At the lower redshift range $0.86 < z < 1.34$, most of works find similar ETG/passive galaxy sizes in clusters and in the field \citeg{Rettura10,Raichoor12,Huertas13a,Kelkar15,Allen15,Saracco17,Marsan19}. \citet{Matharu19} find that cluster quiescent galaxies with $\mathrm{ log} \left(M/M_{\astrosun} \right) \gtrsim 10$ are $0.08 \pm 0.04$ dex ($\sim 20$ \%) more compact than in the field (see also \citealp{Raichoor12}), and are consistent within $1 \sigma$ of the field MSR, which has an intrinsic scatter $\sim 0.13$~dex. 
They used a toy model to show that these galaxies will in part merge with the BCGs and in part be tidally destroyed, and new, larger, galaxies will be accreted into clusters from the field, maintaining a similar MSR in field and clusters in the redshift range $0<z \lesssim 1$. The model considers that galaxies in groups and filaments constantly fall into the cluster haloes over cosmic time \citep{vanDokkumFranx96,Saglia10, Shankar15}. 
Because of this accretion, new members are added to the original passive population. The new passive objects can form either through environmental quenching by ram-pressure \citep{1972ApJ...176....1G}, harassment \citep{Moore96} or strangulation \citep{vandenBosch08}, or they can infall already quenched, by group preprocessing \citep{Fujita04}. In the first case, predominantly late-type disky galaxies are mostly transformed into lenticulars and dwarf ellipticals. In the second case, galaxies preprocessed in the group environment are larger. A part of the size growth in the clusters can be attributed to the addition of group elliptical and lenticular galaxies that mix with the native cluster ETG population and homogenize the size distribution to that of the field (see also \citealp{Matteuzzi22}). This is further compounded by results by \citet{Matharu20}, that can be explained by the accretion of old compact ETGs onto BCGs or their disruption into the intracluster light. The \citet{Matharu19} results at $1<z<1.34$ show larger sizes, and are consistent with this paper.



\subsection{The mass-size relation at $z\sim 2$: Galaxy sizes are larger in clusters than in the field }
Field galaxies at $ 2 \lesssim z \lesssim 3$ \citep{vanderWel14, Patel17, Marsan19} show larger sizes than the extrapolation of the field size evolution at lower redshift (e.g., see the highest redshift field point in Fig~\ref{pic_mei15}). This is explained by the transition from the epoch in which galaxy growth is dominated by gas accretion and the epoch in which minor mergers become dominant \citep{Naab09}.
 On the other end, cluster galaxies are already larger than field galaxies at $z\sim 2$, then grow more slowly than field galaxies, to reach the same average sizes by $z=1$ and then evolve to $z\sim 0$ on average in the same way as field galaxies, mainly because of accretion of field larger galaxies and disruption of the cluster more compact galaxies \citep{Matharu19}.



Our work has highlighted a clear dichotomy in the evolution of the mean passive ETG sizes of similar stellar mass: galaxies in clusters tend to be larger at $z\gtrsim 1-1.5$ than their counterparts in the field, and evolve slower since $z\sim 2$ (see also \citealp{Andreon16}). Passive ETGs in the field are more compact at high redshift, show a faster growth, and eventually show a similar MSR to cluster galaxies at $z\sim 0$, as shown in the works cited above. In other words, passive ETGs of similar stellar mass appear to have a significant environmental (halo) dependence which tends to progressively disappear when approaching the local Universe.

\subsubsection{Model predictions}

This nontrivial evolution in the size evolution is not easily reconciled with theoretical models (see also \citet{Andreon20}). Mergers, especially dry mergers, have traditionally been invoked as the main driver behind the (strong) size evolution of massive galaxies, and in general of all ETGs, as confirmed by a number of cosmological theoretical \citeg{Guo+11, Shankar13,Shankar14a} and numerical \citeg{Naab09, Genel+18, Furlong+17} studies.

\begin{figure}[!]
	\centering
	\includegraphics[width=0.8\hsize]{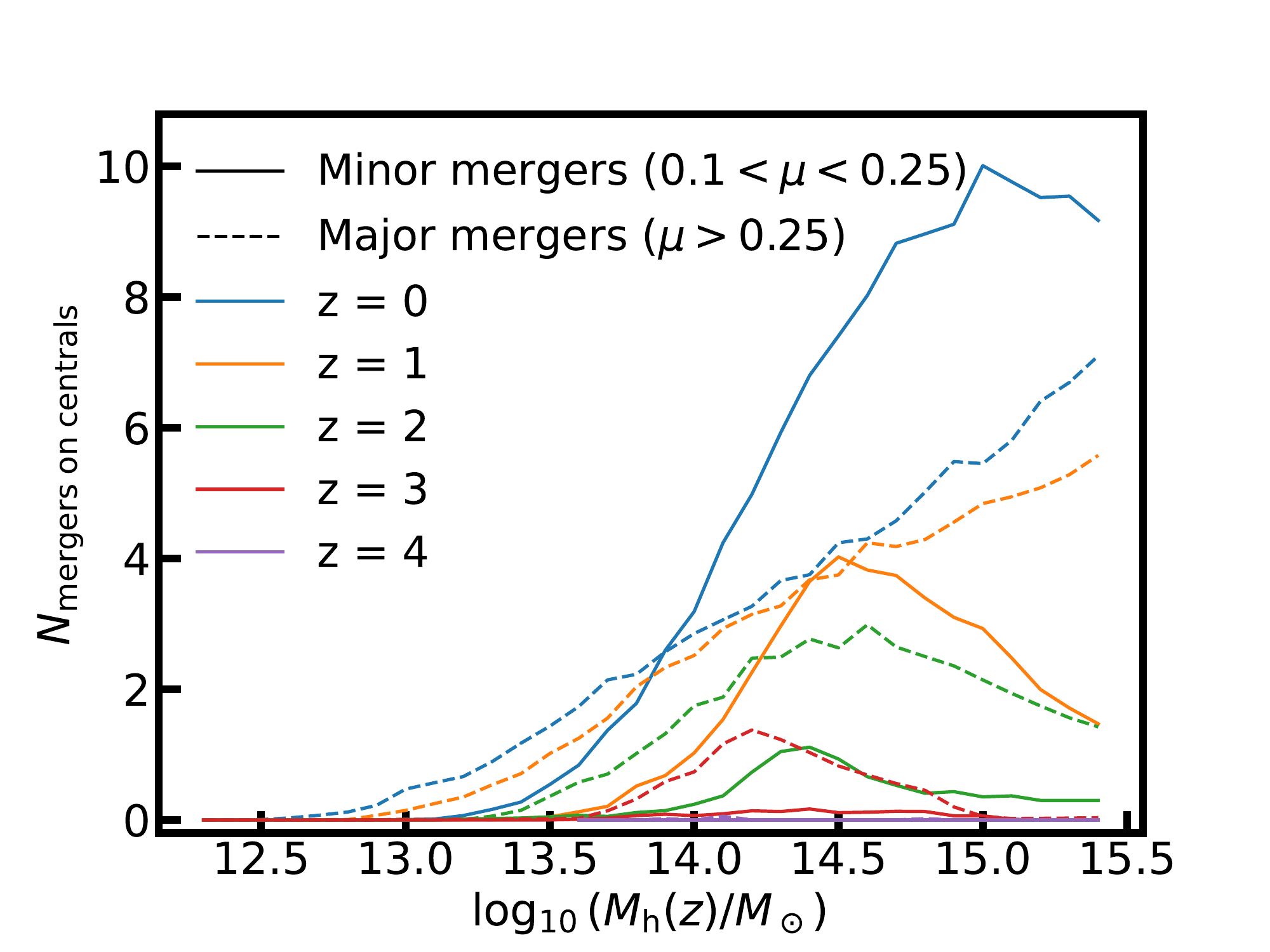}
		\caption{Number of expected major and minor mergers of mass ratios ($\mu$) in the progenitor masses of $\mu>0.25$ and $0.1<\mu<0.25$ (dashed and solid lines, respectively) that a central galaxy of log$\left(M/M_{\astrosun} \right) > 10.5$ undergoes as a function of host halo mass at different redshifts, as labeled. The curves are derived from the DECODE semi-empirical model \citep{Fu22}.} 
		\label{fig:mergers}
\end{figure}

Therefore, one could hypothesize that galaxies in clusters may have undergone a more rapid size increase via mergers before infall into a larger halo. To investigate this hypothesis, Fig~\ref{fig:mergers} shows the predicted cumulative number of expected mergers that a central galaxy of $\mathrm{ log} \left(M/M_{\astrosun} \right) > 10.5$ undergoes as a function of host dark matter halo mass from $z=4$ to present (as labeled) from the DECODE semi-empirical model \citep{Fu22}. DECODE can flexibly compute the (mean) number of mergers of any central galaxy at any given redshift and host halo (cluster or field) without limits of mass or volume resolution and for any given input stellar mass-halo mass relation. Fig~\ref{fig:mergers} shows that the number of both major and minor mergers undergone by central galaxies (dashed and solid lines, with thresholds as labeled) steadily increases as a function of host halo mass at all redshifts. It is interesting to observe that the model predicts that major mergers become more common in halos of mass log$\left(M_h/M_{\astrosun} \right) \sim 14$ at $z \sim 3$, and minor mergers in halos of mass log$\left(M_h/M_{\astrosun} \right) \sim 14.5$ at $z \sim 2$, exactly as it is observed for our CARLA sample in M22.

On the assumption that satellite galaxies with stellar mass $\mathrm{ log} \left(M/M_{\astrosun} \right) > 10.5$ in cluster environments with $\mathrm{ log} \left(M_h/M_{\astrosun} \right)\sim 14$, were, before infall, ``typical'' central galaxies in host haloes of lower mass, then they should have experienced a merger history as the one reported in Fig~\ref{fig:mergers}. In particular, central galaxies with stellar mass $\mathrm{ log} \left(M/M_{\astrosun} \right) > 10.5$ typically reside in host haloes of $\mathrm{ log} \left(M_h/M_{\astrosun} \right) \gtrsim 12.5$ with a weak dependence on redshift \citep[e.g.,][\citealt{Fu22}]{Moster18, Grylls19}. These haloes hardly go through any major or minor merger at $z \gtrsim 2$, according to our predictions shown in Fig~\ref{fig:mergers}. Even allowing for more massive host haloes for more massive galaxies, the models do not predict mergers at $z \gtrsim 2$ up to $\mathrm{ log} \left(M_h/M_{\astrosun} \right)\sim 13.7$, thus disfavoring early size growth via a sequence of minor or major mergers in moderately massive satellite galaxies in clusters, as long as the latter share similar properties and assembly histories to their field counterparts before infall.


Alternatively, progenitor bias could explain the difference between older, more compact and younger, more extended galaxies \citeg{ShankarBernardi09,Lilly16}, or simply the time a galaxy spends in the main sequence could make them larger \citeg{Genel+18}. In our data though, this will imply that the larger cluster galaxies are younger and more extended and spent more time on the main sequence. We cannot measure the age distributions for our galaxies compared to the field, but we expect our cluster galaxies to be older and quenched \citep{Thomas05,Mei22}, and therefore to be smaller and not larger than field galaxies if their size would be different due to the progenitor bias. 

 \begin{figure}[]
	\centering
	\includegraphics[width=0.7\hsize]{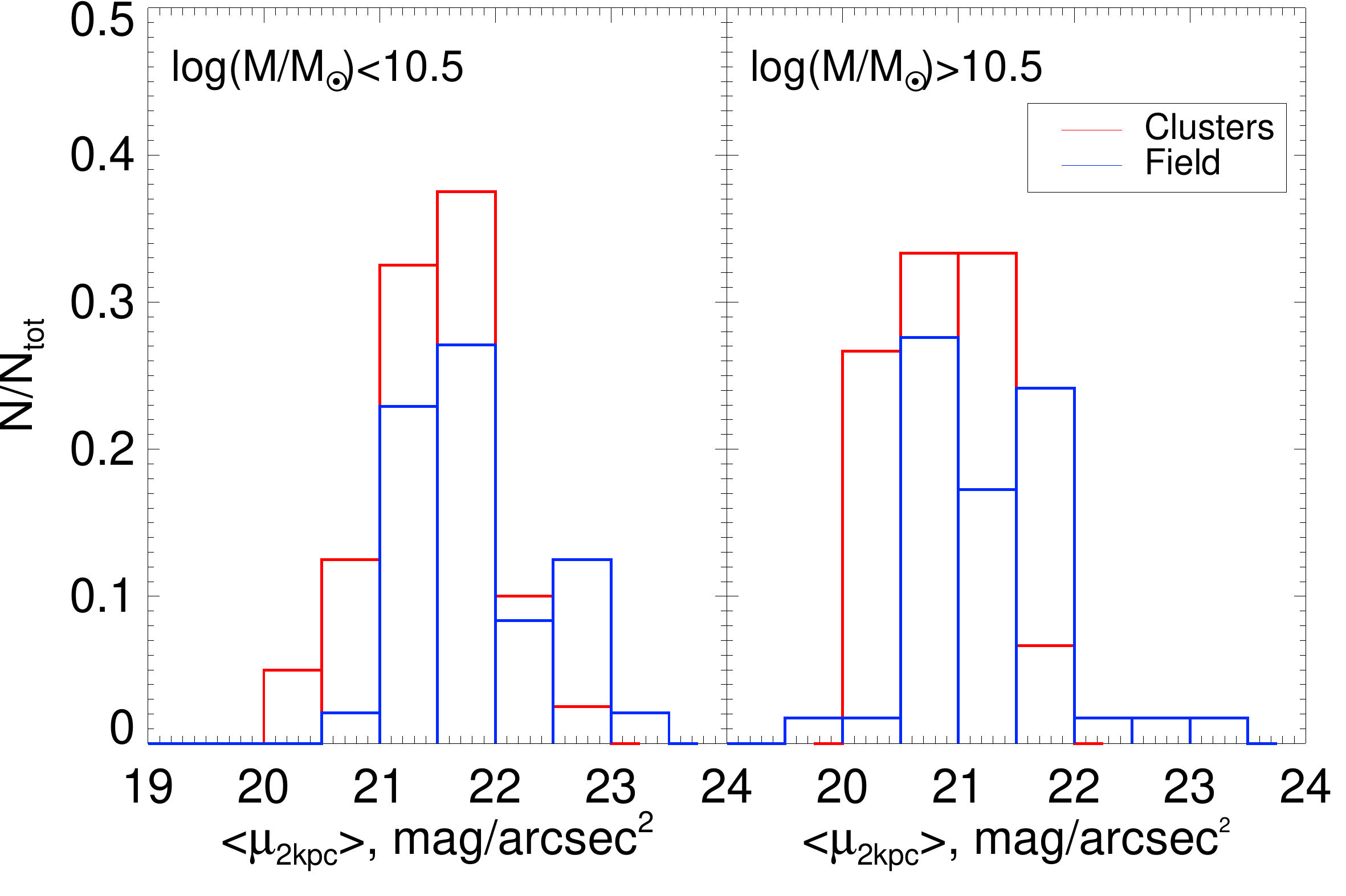}
	\includegraphics[width=0.7\hsize]{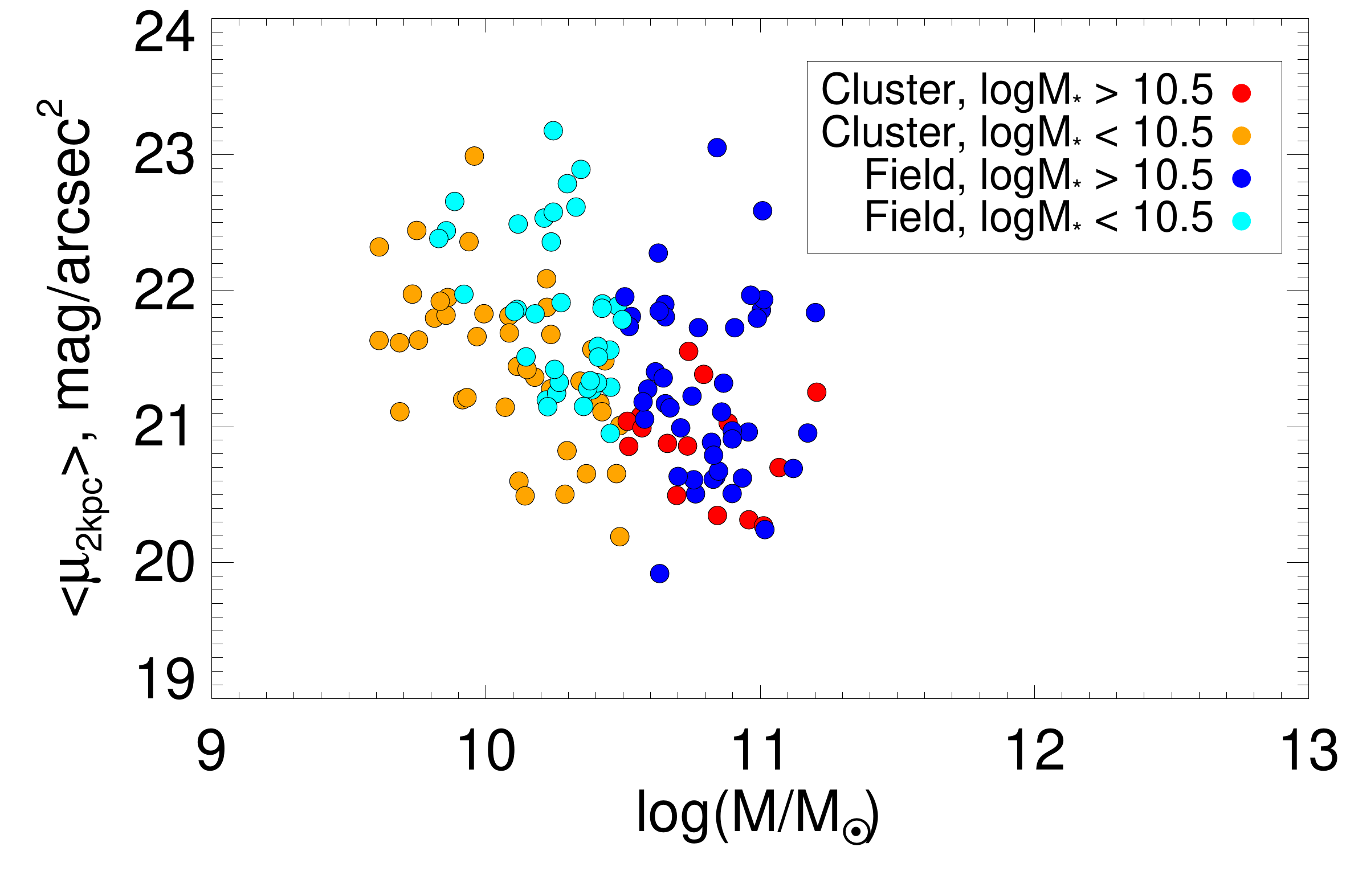}
		\caption{Galaxy average surface brightness in clusters and in the field. Top: Histogram of the galaxy average surface brightness within 2~kpc, normalized to the total number of galaxies in each sample. The continuous red and blue lines show cluster and field galaxies, respectively. Bottom: Dependence of 2~kpc average surface brightness on galaxy mass in mass bins as shown in the labels. More massive galaxies have on average brighter inner regions. In both cases, the figures show that cluster and field galaxies have similar surface brighness distributions.} 
		\label{fig:mu_hist}
\end{figure}

AGN feedback could induce a rapid puffing up of the host galaxy, if a proportionally significant gas mass is expelled from the central regions \citep{Fan08, Fan10}. However, it is not clear why strong AGN feedback should act only in those galaxies destined to become satellites in larger haloes and not in all galaxies of similar stellar mass at a given epoch. In addition, we inspected the central surface brightness within 2 kpc of galaxies of similar stellar mass in the field and in clusters, finding no signs of a reduced central density in cluster galaxies (see Fig.~\ref{fig:mu_hist}), which would be expected if AGN feedback had been expanding the central regions thus decreasing the central densities \citep{Fan10}. 

\citet{Kravtsov13} find evidence of a close linear relation between the effective radii of galaxies and their host haloes of the form $R_e = k\times R_{200c}$, where $R_{200c}$ is the host halo radius\footnote{Defined as the region containing a mass density equal to two hundred times the critical density of the Universe at a given redshift}, and the constant of proportionality $k$ equal to a few percent, depending on the exact definition of host halo mass. This relation is also measured at higher redshifts \citeg{Somerville18}, confirmed in hydrodynamic simulations \citeg{Rohr22}, and it is used in analytic models to show that the size evolution and local size functions of intermediate and massive galaxies can be reproduced \citeg{Stringer14, Zanisi20, Zanisi21a}, along with the environmental halo dependence in the local Universe \citet{Zanisi21b}. 

By using DECODE, we have assigned an effective radius to all centrals and satellite galaxies at different epochs living in the field and in clusters as those in our sample ($13.5 \lesssim \mathrm{ log} \left(M_{h}/M_{\astrosun} \right) \lesssim 14.5$), assuming throughout a constant $R_e= k \times R_{200c}$\footnote{The exact value of the constant $k$ assumed in this exercise is irrelevant as we are only interested in the relative difference between the mean sizes of field and cluster galaxies.}. 
We found that indeed cluster galaxies have a weaker evolution than field galaxies of similar stellar mass, which catch up with their cluster counterparts at $z<0.5$. However, for both predicted cluster and field galaxy radii we find an evolution of the type $H[z]^{2/3}$, that is as $R_{200c}(z)$ \citeg{Stringer14}, which is a weaker evolution than the one observed for ETGs (Figure 6) which is closer to $\sim 1/(1+z)^{\alpha}$, with $\alpha \gtrsim 1$ in the field. This apparent discrepancy could be a sign that the Kravtsov relation is more appropriate to describe the bulk of the population for a given halo/stellar mass, which is represented by star-forming galaxies with $\mathrm{ log} \left(M/M_{\astrosun} \right) \sim 10.5$. We conclude that possibly an imprint in the formation/early evolution of cluster galaxies, as mirrored in the Kravtsov relation, could explain at least in part the systematic difference observed in our sample for cluster and field ETGs, but other factors, such as strong compaction/gas dissipation in field galaxies, followed by a sequence of mergers \citeg{Dekel09,Lapi18} may have also played a significant role in shaping field, but not necessarily cluster, ETGs. 

\subsubsection{Galaxy size proxy and possible bias}

The difference in the average galaxy size evolution might also be due to the choice of the galaxy size proxy that we use. In fact, the half-light radii $R_e$ can be biased by the way the galaxy light is distributed, and, for example, might not correspond to the galaxy sizes measured using mass distribution or other mass proxies \citep{Miller19, Suess19, Miller22}, and might bias the quantification of size evolution if galaxies change concentration while changing size \citep{Andreon20}. 

Mass-to--light ratio gradients in galaxies could have a nonnegligible impact in the calibration and interpretation of the apparent size evolution of galaxies across cosmic time \citep[e.g.,][and references therein]{Hopkins09}. Mass-to-light ratio gradients can be physically caused by gradients in the galaxy stellar populations, with older, more metal-poor, or dustier stellar populations having higher mass-to-light ratios than younger or more metal-rich ones. \citet{Suess19} measure CANDELS galaxy half-mass radii and find that the redshift evolution of galaxy half-mass radii is much slower than that of half-light radii, as also pointed out by \citet{Miller22}. They showed that mass-to-light gradients are stronger for more massive, larger and redder galaxies.
\citet{Bernardi22} also discuss that stellar Initial Mass Function (IMF)-driven gradients might be even stronger than those driven by age and metallicity, and have a larger impact in galaxy size measurement, especially in ETGs. 

While Fig.\ref{fig:mu_hist} shows that our cluster and field samples cover a similar range in galaxy stellar mass, and similar central surface brightness, unfortunately we cannot measure half-mass radii for our cluster sample and study in depth possible bias due to the choice of our mass proxy. However, if galaxies in clusters experienced a different formation and assembly history than their field counterparts, then they could have generated nontrivial mass-to-light or IMF gradients that could induce an apparently weaker size evolution than field galaxies. In particular, our cluster half-light radii seem to better trace mass because they show a slower size evolution that is also observed for field half-mass radii \citep{Suess19}. This suggests that our cluster galaxies might possess less pronounced mass-to-light ratio gradients than galaxies in the field.
 
 \begin{figure}[]
	\centering
	\includegraphics[width=0.7\hsize]{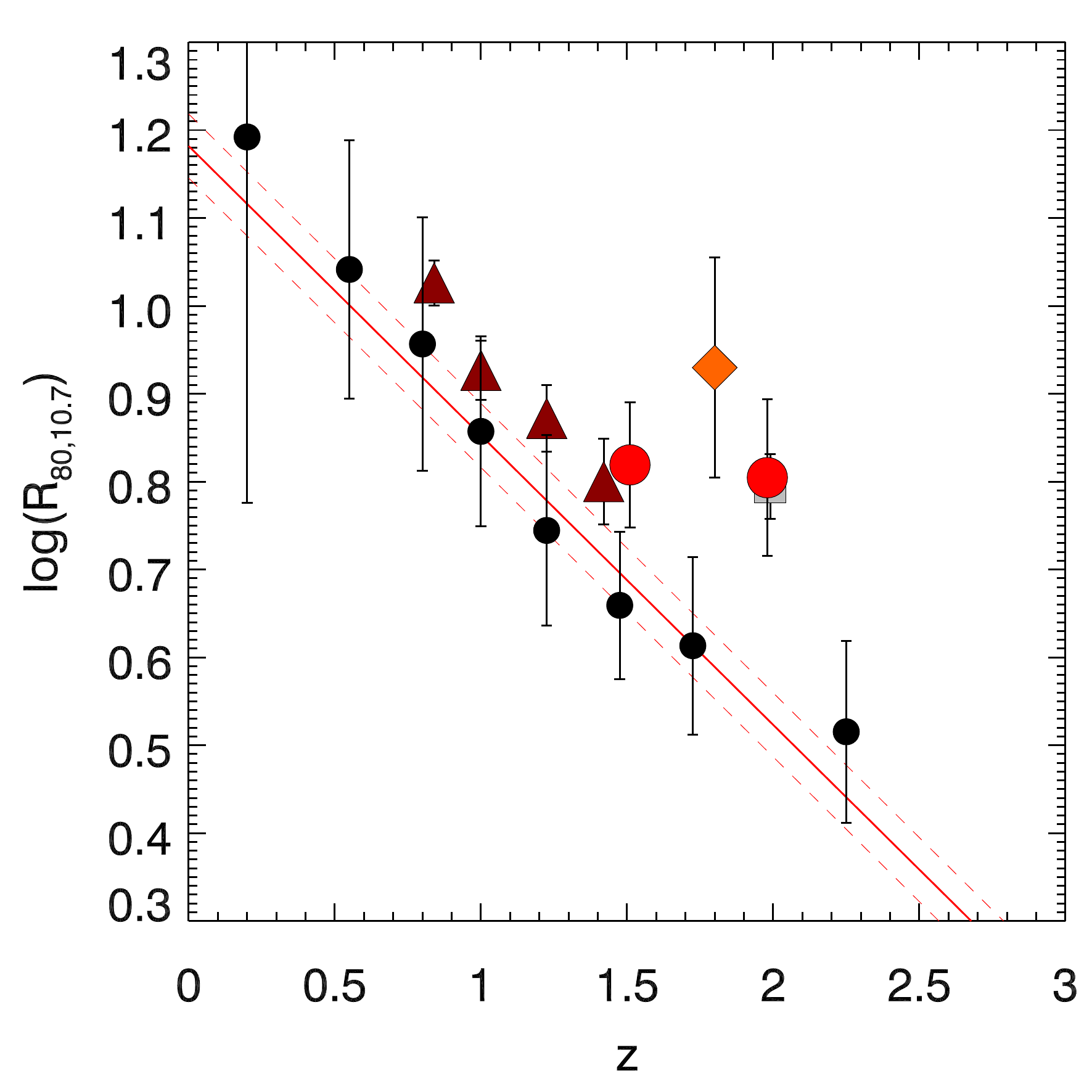}
		\caption{The redshift evolution of the passive ETG size $R_{80}$, normalized at $\mathrm{ log} \left(M/M_{\astrosun} \right) = 10.7$. The symbols are the same as in Fig.~\ref{pic_mei15}.}
		\label{pic_mei15_r80}
\end{figure}

Another way to understand bias due to the choice of using $R_e$ as a size proxy is to measure $R_{80}$ (a radius enclosing $80~\%$ of the galaxy luminosity (see \citealp{Miller19, Mowla19a, Andreon20}), which relates to the galaxy stellar mass in the same way as the dark matter halo mass \citeg{Mowla19}. For a best-fitting S\'{e}rsic profile of each galaxy, the $R_{80}$ value can be calculated analytically \citep{Miller19}.
Figure~\ref{pic_mei15_r80} shows the mass-normalized $R_{80,10.7}$ evolution with redshift. The cluster passive ETG sizes are still $\sim$ 0.3~dex larger ($\gtrsim 2 \sigma$) than the field at $z \gtrsim 1.5$. This is confirmed by \citet{Andreon20} in the redshift range $0.17<z<1.8$ for a sample of cluster ETG more massive than those selected in our work ($\mathrm{ log} \left(M/M_{\astrosun} \right) > 10.7$)). This result also shows a clear dependence of galaxy size on the host halo mass \citep{Kravtsov13}, independently of which light radius proxy is used.


\subsection{BCG sizes}

Our cluster BCGs and the second brightest cluster galaxies lie on the same MSR as the satellite galaxies. Our observations would suggest that the size evolution of BCGs should be similar to other ETG galaxies of similar stellar mass, and quite modest ($\approx 0.15$ dex) at $z<1$ (see also \citealp{Andreon18}). Cosmological models that include size evolution tend to predict, on average, a slightly larger increase ($\approx$0.15-0.3 dex) since $z\sim 1$, depending on the exact model, the selection, and the physical assumptions made during the merger \citeg{2009ApJ...691.1424H, Shankar13, Shankar15, Zoldan19}. Some models also predict that BCGs evolve from normal galaxies at $z=2$, which then become larger due to different merger histories \citep{Zhao17}. This latter work also finds that the most massive cluster galaxy at $z=2$ is a true progenitor of a local BCG less than $50\%$ of the time. Our observations agree with this last scenario, and suggest that BCG sizes evolved as those of the other cluster galaxies at $z \gtrsim 2$.

At lower redshift, observations show different results, even in the local Universe. For example, in the local Universe, \citet{Bernardi09} find that BCGs are systematically larger than satellite galaxies, while \citet{Weinmann09} do not find any size difference between central and satellite ETGs. 



\subsection{Active ETGs that lie on the passive MSR}

We observe active ETGs that follow the passive ETG MSR in four of our five clusters at $z=1.5-2$ (M22). The presence of active ETGs in clusters has been observed at $z\sim 0$ \citeg{Sheen16} and up to $z\sim 2$ \citep{FerrerasSilk2000,Mei06,Mei15,Jaffe11,Mansheim17}. 
 
At $1.35<z<1.65$ the CARLA active ETGs that lie on the passive MSR are $21^{+7}_{-5}$\% of all ETGs, of which $45\pm18$\% lie within $1\sigma$ of the \citet{vanderWel14}'s passive MSR ($64^{+16}_{-20}$\% for 2.5$\sigma$). At $1.65<z<2.05$ active ETG are $59\pm14$\% of all ETG, of which $42\pm17$\% lie within $1\sigma$ of the passive MSR ($58\pm17$\% for 2.5$\sigma$). About half of the active ETGs are mergers or asymmetric (M22). 

The active ETGs that are not mergers and interactions mostly lie on the \citet{vanderWel14}'s passive galaxy MSR as we would expect if their star formation activity did not change their size. Instead, the active ETGs that are mergers or asymmetric lie on the \citet{vanderWel14}'s active galaxy MSR. For these last galaxies, the interactions with other galaxies might have triggered star formation. This might mean that their interactions or asymmetric shapes might have lead to our larger size measurements or that they are misclassified.

In the local Universe ($z<1$), active ETGs are thought to have gone through recent gas-rich minor merger events or interactions with neighboring gas-rich galaxies, and are to become passive when their gas would be exhausted \citeg{Lee06, Huertas10, George15, George17}. Half of our active ETGs are experiencing (M22), and the others very likely have experienced, a recent merger or galaxy interaction, and would most probably quench at a later epoch, thereby increasing the fraction of passive ETGs in the cluster population. Since the higher redshift end of our sample shows high percentage of active ETG that lie within 2.5$\sigma$ of the passive MSR, this suggests that the incidence of recent mergers or neighbor galaxy interactions might have been higher in higher redshift clusters. Since M22 do not observe higher merger fractions in the redshift range that we probe with the CARLA sample, observations at higher redshift are needed to test this hypothesis.


\subsection{The MSR flattening for $\mathrm{log}(M/M_{\odot})\lesssim 10.5$}

Our MSR indicates a tendency to flatten at $\mathrm{ log} \left(M/M_{\astrosun} \right) \lesssim 10.5$. In the range $ 9.6<\mathrm{ log }\left(M/M_{\astrosun} \right) < 10.5$, we measure an average cluster passive ETG size of $\mathrm{log}(R_e/\mathrm{kpc}) = 0.05 \pm 0.22$. This is a trend observed in passive ETGs in clusters and in the field in the local Universe \citeg{Bernardi14, Nedkova21} and at $z \lesssim 2$ \citeg{Nedkova21} for galaxies in the mass range $ 7 \lesssim \mathrm{log}(M/M_{\odot})\lesssim 10.5$. It is predicted in semi-analytical models \citeg{Shankar13}, where it occurs at the transitional mass $\mathrm{log}(M/M_{\odot})\sim 10.5$, below which galaxy growth is dominated by both disk instabilities and mergers, and above which galaxy growth is dominated by minor mergers. Our average cluster passive ETG size is $\sim 0.2$~dex smaller than the average size from \citet{Nedkova21}, and consistent within $\sim 1\sigma$. This suggests that the low mass end of the MSR does not evolve much from $z\sim2$ to present.



\section{Summary} \label{sec:disc}

We studied the MSR of galaxies in a sample of 15 spectroscopically confirmed clusters from the CARLA survey \citep{Wylezalek13,Wylezalek14, Noirot18}. Our cluster total stellar mass spans ithe range $11.3<\log(M^c_*/M_{\astrosun})<12.6$, which corresponds to an approximate halo mass in the range $13.5 \lesssim \log(M^c_h/M_{\astrosun}) \lesssim 14.5$ (M22). 

\vskip 0.2cm

Our main results are:

\begin{itemize}

\item Cluster LTGs at $1.4<z<2.8$ lie on the same MSR as active field galaxies from CANDELS \citep{vanderWel14}.

\vskip 0.2cm

\item Cluster ETGs at $1.4<z<2.8$ show sizes that are $\sim 0.3$~dex ($\gtrsim~3\sigma$) systematically larger than passive field galaxies from CANDELS \citep{vanderWel14}. The evolution of cluster passive ETG sizes is slower at $1<z<2$ when compared to the field. We fit the average evolution for the mass-normalized radius as:
\begin{equation}
{\rm log}(R_{10.7}/\mathrm{kpc})=(-0.16 \pm 0.02) \ (z-1)+(0.44 \pm 0.01),
\end{equation}
compared to the evolution in the field from \citet{vanderWel14}: 
\begin{equation}
{\rm log}(R_{10.7}/\mathrm{kpc})=(-0.28 \pm 0.04)\ (z-1)+(0.33 \pm 0.02).
\end{equation}

\item BCGs lie on the same MSR as the satellites. 
\vskip 0.2cm
\item Half of the active ETGs follow the field passive galaxy MSR, and the other half the active galaxy MSR. 
\vskip 0.2cm

\item In the range $ 9.6<\mathrm{ log }\left(M/M_{\astrosun} \right) < 10.5$, our passive ETG MSR is consistent with flattening, with an average $\mathrm{log}(R_e/\mathrm{kpc}) = 0.05 \pm 0.22$. It is $\sim 0.2$~dex smaller than the field studies in the similar mass range at $z=0-2$ \citep{Nedkova21}, but the two results are consistent within $1\sigma$.
\vskip 0.2cm
\item We do not observe a large population of compact galaxies (only one), in contrast with field studies at these redshifts \citeg{Barro13}, and studies that found high percentages of compact post-starburst (\citealp{Maltby18,Socolovsky19, Matharu20, Wilkinson21}).
\end{itemize}
\vskip 0.2cm

In conclusion, the systematic difference in size that we observe between cluster and field passive ETG galaxies could most probably be explained by early-epoch differences in the formation and early evolution of galaxies in haloes of different mass, as predicted by models \citep{Kravtsov13}. However, other physical mechanisms, such as strong compaction/gas dissipation in field galaxies, followed by a sequence of mergers \citeg{2009ApJ...691.1424H,Lapi18} could play a role in field ETG galaxy evolution, but not necessarily in the evolution of cluster galaxies. The low-mass end of the MSR ($ 9.6<\mathrm{ log }\left(M/M_{\astrosun} \right) < 10.5$) did not evolve much from $z\sim2$ to the present and does not show significant environmental dependence. This suggests that the physical mechanisms that govern these low-mass galaxies are similar in clusters and in the field. We also find that the BCGs lie on the same MSR as the other cluster galaxies, implying that their size evolution is not very different from the other cluster galaxies at $z \gtrsim 2$. 

 Our active ETGs that are not mergers and interactions mostly lie on the \citet{vanderWel14}'s passive galaxy MSR as we would expect if their star formation activity did not change their size. Instead, the active ETGs that lie on the \citet{vanderWel14}'s active galaxy MSR are mostly mergers and asymmetric, where the interactions with other galaxies might have triggered star formation. This might mean that their interactions or asymmetric shapes might have lead to our larger size measurements or that they were misclassified. These ETGs would most probably quench at a later epoch thereby increasing the fraction of passive ETGs in the cluster. Our sample also shows a lack of compact galaxies. This implies that the galaxies in our clusters are not observed at an epoch close to their compaction \citeg{Dekel09, Barro13}, which might have happened at higher redshift in the rich cluster environments \citeg{Lustig21}.




\newpage

\begin{acknowledgements}

This work is based on observations made with the NASA/ESA {\it Hubble Space Telescope}, obtained at the Space Telescope Science Institute, which is operated by the Association of Universities for Research in Astronomy Inc., under NASA contract NAS 5-26555. These observations are associated with program GO-13740. Support for program GO-13740 was provided by NASA through a grant from the Space Telescope Science Institute, which is operated by the Association of Universities for Research in Astronomy Inc., under NASA contract NAS 5-26555. HF and FS acknowledge support from the European Union's Horizon 2020 research and innovation programme under the Marie Skłodowska-Curie grant agreement No. 860744. This work is based on observations made with the {\it Spitzer Space Telescope}, which is operated by the Jet Propulsion Laboratory, California Institute of Technology, under a contract with NASA. We thank Ignacio Trujillo, Arjen van der Wel, Igor Chilingarian, and Francoise Combes for useful comments. We thank Leo Girardi for his help with the TRILEGAL model. We thank Universit\'e Paris Cité, which founded AA' s Ph.D. research. SM thanks Jet Propulsion Laboratory, California Institute of Technology, for hosting her in the context of this project.
The work of DS was carried out at the Jet Propulsion Laboratory, California Institute of Technology, under a contract with NASA. GN acknowledges funding support from the Natural Sciences and Engineering Research Council (NSERC) of Canada through a Discovery Grant and Discovery Accelerator Supplement, and from the Canadian Space Agency through grant 18JWST-GTO1. NAH thanks the Science and Technology Facilities Council, UK, consolidated grant ST/T000171/1. 
This work was supported by the French Space Agency (CNES). 
We thank the anonymous referee for her/his careful reading of the manuscript and useful suggestions that helped to improve the paper.

\end{acknowledgements}

\bibliographystyle{aa}
\bibliography{CARLA_mass-size}

\begin{thebibliography}{157}
\expandafter\ifx\csname natexlab\endcsname\relax\def\natexlab#1{#1}\fi

\bibitem[{{Allen} {et~al.}(2015){Allen}, {Kacprzak}, {Spitler}, {Glazebrook},
  {Labb{\'e}}, {Tran}, {Straatman}, {Nanayakkara}, {Brammer}, {Quadri},
  {Cowley}, {Monson}, {Papovich}, {Persson}, {Rees}, {Tilvi}, \&
  {Tomczak}}]{Allen15}
{Allen}, R.~J., {Kacprzak}, G.~G., {Spitler}, L.~R., {et~al.} 2015, \apj, 806,
  3

\bibitem[{{Andreon}(2018)}]{Andreon18}
{Andreon}, S. 2018, \aap, 617, A53

\bibitem[{{Andreon}(2020)}]{Andreon20}
{Andreon}, S. 2020, \aap, 640, A34

\bibitem[{{Andreon} {et~al.}(2016){Andreon}, {Dong}, \& {Raichoor}}]{Andreon16}
{Andreon}, S., {Dong}, H., \& {Raichoor}, A. 2016, \aap, 593, A2

\bibitem[{{Barden} {et~al.}(2012){Barden}, {H{\"a}u{\ss}ler}, {Peng},
  {McIntosh}, \& {Guo}}]{barden12}
{Barden}, M., {H{\"a}u{\ss}ler}, B., {Peng}, C.~Y., {McIntosh}, D.~H., \&
  {Guo}, Y. 2012, \mnras, 422, 449

\bibitem[{{Barro} {et~al.}(2013){Barro}, {Faber}, {P{\'e}rez-Gonz{\'a}lez},
  {Koo}, {Williams}, {Kocevski}, {Trump}, {Mozena}, {McGrath}, {van der Wel},
  {Wuyts}, {Bell}, {Croton}, {Ceverino}, {Dekel}, {Ashby}, {Cheung},
  {Ferguson}, {Fontana}, {Fang}, {Giavalisco}, {Grogin}, {Guo}, {Hathi},
  {Hopkins}, {Huang}, {Koekemoer}, {Kartaltepe}, {Lee}, {Newman}, {Porter},
  {Primack}, {Ryan}, {Rosario}, {Somerville}, {Salvato}, \& {Hsu}}]{Barro13}
{Barro}, G., {Faber}, S.~M., {P{\'e}rez-Gonz{\'a}lez}, P.~G., {et~al.} 2013,
  \apj, 765, 104

\bibitem[{{Belli} {et~al.}(2014){Belli}, {Newman}, \& {Ellis}}]{Belli14}
{Belli}, S., {Newman}, A.~B., \& {Ellis}, R.~S. 2014, \apj, 783, 117

\bibitem[{{Bernardi}(2009)}]{Bernardi09}
{Bernardi}, M. 2009, \mnras, 395, 1491

\bibitem[{{Bernardi} {et~al.}(2013){Bernardi}, {Meert}, {Sheth}, {Vikram},
  {Huertas-Company}, {Mei}, \& {Shankar}}]{Bernardi13}
{Bernardi}, M., {Meert}, A., {Sheth}, R.~K., {et~al.} 2013, \mnras, 436, 697

\bibitem[{{Bernardi} {et~al.}(2014){Bernardi}, {Meert}, {Vikram},
  {Huertas-Company}, {Mei}, {Shankar}, \& {Sheth}}]{Bernardi14}
{Bernardi}, M., {Meert}, A., {Vikram}, V., {et~al.} 2014, \mnras, 443, 874

\bibitem[{{Bernardi} {et~al.}(2011{\natexlab{a}}){Bernardi}, {Roche},
  {Shankar}, \& {Sheth}}]{Bernardi11a}
{Bernardi}, M., {Roche}, N., {Shankar}, F., \& {Sheth}, R.~K.
  2011{\natexlab{a}}, \mnras, 412, 684

\bibitem[{{Bernardi} {et~al.}(2011{\natexlab{b}}){Bernardi}, {Roche},
  {Shankar}, \& {Sheth}}]{Bernardi11b}
{Bernardi}, M., {Roche}, N., {Shankar}, F., \& {Sheth}, R.~K.
  2011{\natexlab{b}}, \mnras, 412, L6

\bibitem[{{Bernardi} {et~al.}(2022){Bernardi}, {Sheth}, {S{\'a}nchez},
  {Margalef-Bentabol}, {Bizyaev}, \& {Lane}}]{Bernardi22}
{Bernardi}, M., {Sheth}, R.~K., {S{\'a}nchez}, H.~D., {et~al.} 2022, \mnras
  [\eprint[arXiv]{2201.07810}]

\bibitem[{{Bertin} \& {Arnouts}(1996)}]{sextractor}
{Bertin}, E. \& {Arnouts}, S. 1996, \aaps, 117, 393

\bibitem[{{Boselli} \& {Gavazzi}(2006)}]{2006PASP..118..517B}
{Boselli}, A. \& {Gavazzi}, G. 2006, \pasp, 118, 517

\bibitem[{{Cappellari}(2013)}]{Cappellari13}
{Cappellari}, M. 2013, \apjl, 778, L2

\bibitem[{{Carollo} {et~al.}(2013){Carollo}, {Bschorr}, {Renzini}, {Lilly},
  {Capak}, {Cibinel}, {Ilbert}, {Onodera}, {Scoville}, {Cameron}, {Mobasher},
  {Sanders}, \& {Taniguchi}}]{Carollo13}
{Carollo}, C.~M., {Bschorr}, T.~J., {Renzini}, A., {et~al.} 2013, \apj, 773,
  112

\bibitem[{{Chabrier}(2003)}]{2003PASP..115..763C}
{Chabrier}, G. 2003, \pasp, 115, 763

\bibitem[{{Chan} {et~al.}(2018){Chan}, {Beifiori}, {Saglia}, {Mendel}, {Stott},
  {Bender}, {Galametz}, {Wilman}, {Cappellari}, {Davies}, {Houghton},
  {Prichard}, {Lewis}, {Sharples}, \& {Wegner}}]{Chan18}
{Chan}, J. C.~C., {Beifiori}, A., {Saglia}, R.~P., {et~al.} 2018, \apj, 856, 8

\bibitem[{{Chiang} {et~al.}(2013){Chiang}, {Overzier}, \&
  {Gebhardt}}]{Chiang13}
{Chiang}, Y.-K., {Overzier}, R., \& {Gebhardt}, K. 2013, \apj, 779, 127

\bibitem[{{Cooke} {et~al.}(2015){Cooke}, {Hatch}, {Rettura}, {Wylezalek},
  {Galametz}, {Stern}, {Brodwin}, {Muldrew}, {Almaini}, {Conselice},
  {Eisenhardt}, {Hartley}, {Jarvis}, {Seymour}, \& {Stanford}}]{Cooke15}
{Cooke}, E.~A., {Hatch}, N.~A., {Rettura}, A., {et~al.} 2015, \mnras, 452, 2318

\bibitem[{{Dekel} {et~al.}(2009){Dekel}, {Birnboim}, {Engel}, {Freundlich},
  {Goerdt}, {Mumcuoglu}, {Neistein}, {Pichon}, {Teyssier}, \&
  {Zinger}}]{Dekel09}
{Dekel}, A., {Birnboim}, Y., {Engel}, G., {et~al.} 2009, \nat, 457, 451

\bibitem[{{Delaye} {et~al.}(2014){Delaye}, {Huertas-Company}, {Mei}, {Lidman},
  {Licitra}, {Newman}, {Raichoor}, {Shankar}, {Barrientos}, {Bernardi},
  {Cerulo}, {Couch}, {Demarco}, {Mu{\~n}oz}, {S{\'a}nchez-Janssen}, \&
  {Tanaka}}]{Delaye14}
{Delaye}, L., {Huertas-Company}, M., {Mei}, S., {et~al.} 2014, \mnras, 441, 203

\bibitem[{{Demers} {et~al.}(2019){Demers}, {Parker}, \& {Roberts}}]{Demers19}
{Demers}, M.~L., {Parker}, L.~C., \& {Roberts}, I.~D. 2019, \mnras, 489, 2216

\bibitem[{{Dimauro} {et~al.}(2019){Dimauro}, {Huertas-Company}, {Daddi},
  {P{\'e}rez-Gonz{\'a}lez}, {Bernardi}, {Caro}, {Cattaneo}, {H{\"a}u{\ss}ler},
  {Kuchner}, {Shankar}, {Barro}, {Buitrago}, {Faber}, {Kocevski}, {Koekemoer},
  {Koo}, {Mei}, {Peletier}, {Primack}, {Rodriguez-Puebla}, {Salvato}, \&
  {Tuccillo}}]{Dimauro19}
{Dimauro}, P., {Huertas-Company}, M., {Daddi}, E., {et~al.} 2019, \mnras, 489,
  4135

\bibitem[{{Dutton} {et~al.}(2011){Dutton}, {van den Bosch}, {Faber}, {Simard},
  {Kassin}, {Koo}, {Bundy}, {Huang}, {Weiner}, {Cooper}, {Newman}, {Mozena}, \&
  {Koekemoer}}]{Dutton11}
{Dutton}, A.~A., {van den Bosch}, F.~C., {Faber}, S.~M., {et~al.} 2011, \mnras,
  410, 1660

\bibitem[{{Erben} {et~al.}(2005){Erben}, {Schirmer}, {Dietrich}, {Cordes},
  {Haberzettl}, {Hetterscheidt}, {Hildebrandt}, {Schmithuesen}, {Schneider},
  {Simon}, {Deul}, {Hook}, {Kaiser}, {Radovich}, {Benoist}, {Nonino}, {Olsen},
  {Prandoni}, {Wichmann}, {Zaggia}, {Bomans}, {Dettmar}, \&
  {Miralles}}]{Erben05}
{Erben}, T., {Schirmer}, M., {Dietrich}, J.~P., {et~al.} 2005, Astronomische
  Nachrichten, 326, 432

\bibitem[{{Erwin} {et~al.}(2012){Erwin}, {Guti{\'e}rrez}, \&
  {Beckman}}]{Erwin12}
{Erwin}, P., {Guti{\'e}rrez}, L., \& {Beckman}, J.~E. 2012, \apjl, 744, L11

\bibitem[{{Fan} {et~al.}(2010){Fan}, {Lapi}, {Bressan}, {Bernardi}, {De Zotti},
  \& {Danese}}]{Fan10}
{Fan}, L., {Lapi}, A., {Bressan}, A., {et~al.} 2010, \apj, 718, 1460

\bibitem[{{Fan} {et~al.}(2008){Fan}, {Lapi}, {De Zotti}, \& {Danese}}]{Fan08}
{Fan}, L., {Lapi}, A., {De Zotti}, G., \& {Danese}, L. 2008, \apjl, 689, L101

\bibitem[{{Fang} {et~al.}(2018){Fang}, {Faber}, {Koo}, {Rodr{\'\i}guez-Puebla},
  {Guo}, {Barro}, {Behroozi}, {Brammer}, {Chen}, {Dekel}, {Ferguson},
  {Gawiser}, {Giavalisco}, {Kartaltepe}, {Kocevski}, {Koekemoer}, {McGrath},
  {McIntosh}, {Newman}, {Pacifici}, {Pandya}, {P{\'e}rez-Gonz{\'a}lez},
  {Primack}, {Salmon}, {Trump}, {Weiner}, {Willner}, {Acquaviva}, {Dahlen},
  {Finkelstein}, {Finlator}, {Fontana}, {Galametz}, {Grogin}, {Gruetzbauch},
  {Johnson}, {Mobasher}, {Papovich}, {Pforr}, {Salvato}, {Santini}, {van der
  Wel}, {Wiklind}, \& {Wuyts}}]{Fang18}
{Fang}, J.~J., {Faber}, S.~M., {Koo}, D.~C., {et~al.} 2018, \apj, 858, 100

\bibitem[{{Fern{\'a}ndez Lorenzo} {et~al.}(2013){Fern{\'a}ndez Lorenzo},
  {Sulentic}, {Verdes-Montenegro}, \&
  {Argudo-Fern{\'a}ndez}}]{FernandezLorenzo13}
{Fern{\'a}ndez Lorenzo}, M., {Sulentic}, J., {Verdes-Montenegro}, L., \&
  {Argudo-Fern{\'a}ndez}, M. 2013, \mnras, 434, 325

\bibitem[{{Ferreras} \& {Silk}(2000)}]{FerrerasSilk2000}
{Ferreras}, I. \& {Silk}, J. 2000, \apjl, 541, L37

\bibitem[{{Fu} {et~al.}(2022){Fu}, {Shankar}, {Ayromlou}, {Dickson},
  {Koutsouridou}, {Rosas-Guevara}, {Marsden}, {Brocklebank}, {Bernardi},
  {Shiamtanis}, {Williams}, {Zanisi}, {Allevato}, {Boco}, {Bonoli}, {Cattaneo},
  {Dimauro}, {Jiang}, {Lapi}, {Menci}, {Petropoulou}, \& {Villforth}}]{Fu22}
{Fu}, H., {Shankar}, F., {Ayromlou}, M., {et~al.} 2022, \mnras, 516, 3206

\bibitem[{{Fujita}(2004)}]{Fujita04}
{Fujita}, Y. 2004, \pasj, 56, 29

\bibitem[{{Furlong} {et~al.}(2017){Furlong}, {Bower}, {Crain}, {Schaye},
  {Theuns}, {Trayford}, {Qu}, {Schaller}, {Berthet}, \& {Helly}}]{Furlong+17}
{Furlong}, M., {Bower}, R.~G., {Crain}, R.~A., {et~al.} 2017, \mnras, 465, 722

\bibitem[{{Gadotti}(2009)}]{Gadotti09}
{Gadotti}, D.~A. 2009, \mnras, 393, 1531

\bibitem[{{Gaia Collaboration} {et~al.}(2021){Gaia Collaboration}, {Brown},
  {Vallenari}, {Prusti}, {de Bruijne}, {Babusiaux}, {Biermann}, {Creevey},
  {Evans}, {Eyer}, {Hutton}, {Jansen}, {Jordi}, {Klioner}, {Lammers},
  {Lindegren}, {Luri}, {Mignard}, {Panem}, {Pourbaix}, {Randich}, {Sartoretti},
  {Soubiran}, {Walton}, {Arenou}, {Bailer-Jones}, {Bastian}, {Cropper},
  {Drimmel}, {Katz}, {Lattanzi}, {van Leeuwen}, {Bakker}, {Cacciari},
  {Casta{\~n}eda}, {De Angeli}, {Ducourant}, {Fabricius}, {Fouesneau},
  {Fr{\'e}mat}, {Guerra}, {Guerrier}, {Guiraud}, {Jean-Antoine Piccolo},
  {Masana}, {Messineo}, {Mowlavi}, {Nicolas}, {Nienartowicz}, {Pailler},
  {Panuzzo}, {Riclet}, {Roux}, {Seabroke}, {Sordo}, {Tanga}, {Th{\'e}venin},
  {Gracia-Abril}, {Portell}, {Teyssier}, {Altmann}, {Andrae}, {Bellas-Velidis},
  {Benson}, {Berthier}, {Blomme}, {Brugaletta}, {Burgess}, {Busso}, {Carry},
  {Cellino}, {Cheek}, {Clementini}, {Damerdji}, {Davidson}, {Delchambre},
  {Dell'Oro}, {Fern{\'a}ndez-Hern{\'a}ndez}, {Galluccio}, {Garc{\'\i}a-Lario},
  {Garcia-Reinaldos}, {Gonz{\'a}lez-N{\'u}{\~n}ez}, {Gosset}, {Haigron},
  {Halbwachs}, {Hambly}, {Harrison}, {Hatzidimitriou}, {Heiter},
  {Hern{\'a}ndez}, {Hestroffer}, {Hodgkin}, {Holl}, {Jan{\ss}en}, {Jevardat de
  Fombelle}, {Jordan}, {Krone-Martins}, {Lanzafame}, {L{\"o}ffler}, {Lorca},
  {Manteiga}, {Marchal}, {Marrese}, {Moitinho}, {Mora}, {Muinonen}, {Osborne},
  {Pancino}, {Pauwels}, {Petit}, {Recio-Blanco}, {Richards}, {Riello},
  {Rimoldini}, {Robin}, {Roegiers}, {Rybizki}, {Sarro}, {Siopis}, {Smith},
  {Sozzetti}, {Ulla}, {Utrilla}, {van Leeuwen}, {van Reeven}, {Abbas}, {Abreu
  Aramburu}, {Accart}, {Aerts}, {Aguado}, {Ajaj}, {Altavilla}, {{\'A}lvarez},
  {{\'A}lvarez Cid-Fuentes}, {Alves}, {Anderson}, {Anglada Varela}, {Antoja},
  {Audard}, {Baines}, {Baker}, {Balaguer-N{\'u}{\~n}ez}, {Balbinot}, {Balog},
  {Barache}, {Barbato}, {Barros}, {Barstow}, {Bartolom{\'e}}, {Bassilana},
  {Bauchet}, {Baudesson-Stella}, {Becciani}, {Bellazzini}, {Bernet}, {Bertone},
  {Bianchi}, {Blanco-Cuaresma}, {Boch}, {Bombrun}, {Bossini}, {Bouquillon},
  {Bragaglia}, {Bramante}, {Breedt}, {Bressan}, {Brouillet}, {Bucciarelli},
  {Burlacu}, {Busonero}, {Butkevich}, {Buzzi}, {Caffau}, {Cancelliere},
  {C{\'a}novas}, {Cantat-Gaudin}, {Carballo}, {Carlucci}, {Carnerero},
  {Carrasco}, {Casamiquela}, {Castellani}, {Castro-Ginard}, {Castro Sampol},
  {Chaoul}, {Charlot}, {Chemin}, {Chiavassa}, {Cioni}, {Comoretto}, {Cooper},
  {Cornez}, {Cowell}, {Crifo}, {Crosta}, {Crowley}, {Dafonte}, {Dapergolas},
  {David}, {David}, {de Laverny}, {De Luise}, {De March}, {De Ridder}, {de
  Souza}, {de Teodoro}, {de Torres}, {del Peloso}, {del Pozo}, {Delbo},
  {Delgado}, {Delgado}, {Delisle}, {Di Matteo}, {Diakite}, {Diener},
  {Distefano}, {Dolding}, {Eappachen}, {Edvardsson}, {Enke}, {Esquej}, {Fabre},
  {Fabrizio}, {Faigler}, {Fedorets}, {Fernique}, {Fienga}, {Figueras},
  {Fouron}, {Fragkoudi}, {Fraile}, {Franke}, {Gai}, {Garabato},
  {Garcia-Gutierrez}, {Garc{\'\i}a-Torres}, {Garofalo}, {Gavras}, {Gerlach},
  {Geyer}, {Giacobbe}, {Gilmore}, {Girona}, {Giuffrida}, {Gomel}, {Gomez},
  {Gonzalez-Santamaria}, {Gonz{\'a}lez-Vidal}, {Granvik},
  {Guti{\'e}rrez-S{\'a}nchez}, {Guy}, {Hauser}, {Haywood}, {Helmi}, {Hidalgo},
  {Hilger}, {H{\l}adczuk}, {Hobbs}, {Holland}, {Huckle}, {Jasniewicz},
  {Jonker}, {Juaristi Campillo}, {Julbe}, {Karbevska}, {Kervella}, {Khanna},
  {Kochoska}, {Kontizas}, {Kordopatis}, {Korn}, {Kostrzewa-Rutkowska},
  {Kruszy{\'n}ska}, {Lambert}, {Lanza}, {Lasne}, {Le Campion}, {Le Fustec},
  {Lebreton}, {Lebzelter}, {Leccia}, {Leclerc}, {Lecoeur-Taibi}, {Liao},
  {Licata}, {Lindstr{\o}m}, {Lister}, {Livanou}, {Lobel}, {Madrero Pardo},
  {Managau}, {Mann}, {Marchant}, {Marconi}, {Marcos Santos}, {Marinoni},
  {Marocco}, {Marshall}, {Martin Polo}, {Mart{\'\i}n-Fleitas}, {Masip},
  {Massari}, {Mastrobuono-Battisti}, {Mazeh}, {McMillan}, {Messina},
  {Michalik}, {Millar}, {Mints}, {Molina}, {Molinaro}, {Moln{\'a}r},
  {Montegriffo}, {Mor}, {Morbidelli}, {Morel}, {Morris}, {Mulone}, {Munoz},
  {Muraveva}, {Murphy}, {Musella}, {Noval}, {Ord{\'e}novic}, {Orr{\`u}},
  {Osinde}, {Pagani}, {Pagano}, {Palaversa}, {Palicio}, {Panahi}, {Pawlak},
  {Pe{\~n}alosa Esteller}, {Penttil{\"a}}, {Piersimoni}, {Pineau}, {Plachy},
  {Plum}, {Poggio}, {Poretti}, {Poujoulet}, {Pr{\v{s}}a}, {Pulone}, {Racero},
  {Ragaini}, {Rainer}, {Raiteri}, {Rambaux}, {Ramos}, {Ramos-Lerate}, {Re
  Fiorentin}, {Regibo}, {Reyl{\'e}}, {Ripepi}, {Riva}, {Rixon}, {Robichon},
  {Robin}, {Roelens}, {Rohrbasser}, {Romero-G{\'o}mez}, {Rowell}, {Royer},
  {Rybicki}, {Sadowski}, {Sagrist{\`a} Sell{\'e}s}, {Sahlmann}, {Salgado},
  {Salguero}, {Samaras}, {Sanchez Gimenez}, {Sanna}, {Santove{\~n}a},
  {Sarasso}, {Schultheis}, {Sciacca}, {Segol}, {Segovia}, {S{\'e}gransan},
  {Semeux}, {Shahaf}, {Siddiqui}, {Siebert}, {Siltala}, {Slezak}, {Smart},
  {Solano}, {Solitro}, {Souami}, {Souchay}, {Spagna}, {Spoto}, {Steele},
  {Steidelm{\"u}ller}, {Stephenson}, {S{\"u}veges}, {Szabados}, {Szegedi-Elek},
  {Taris}, {Tauran}, {Taylor}, {Teixeira}, {Thuillot}, {Tonello}, {Torra},
  {Torra}, {Turon}, {Unger}, {Vaillant}, {van Dillen}, {Vanel}, {Vecchiato},
  {Viala}, {Vicente}, {Voutsinas}, {Weiler}, {Wevers}, {Wyrzykowski}, {Yoldas},
  {Yvard}, {Zhao}, {Zorec}, {Zucker}, {Zurbach}, \& {Zwitter}}]{GaiaEDR3}
{Gaia Collaboration}, {Brown}, A.~G.~A., {Vallenari}, A., {et~al.} 2021, \aap,
  649, A1

\bibitem[{{Galametz} {et~al.}(2013){Galametz}, {Grazian}, {Fontana},
  {Ferguson}, {Ashby}, {Barro}, {Castellano}, {Dahlen}, {Donley}, {Faber},
  {Grogin}, {Guo}, {Huang}, {Kocevski}, {Koekemoer}, {Lee}, {McGrath}, {Peth},
  {Willner}, {Almaini}, {Cooper}, {Cooray}, {Conselice}, {Dickinson}, {Dunlop},
  {Fazio}, {Foucaud}, {Gardner}, {Giavalisco}, {Hathi}, {Hartley}, {Koo},
  {Lai}, {de Mello}, {McLure}, {Lucas}, {Paris}, {Pentericci}, {Santini},
  {Simpson}, {Sommariva}, {Targett}, {Weiner}, {Wuyts}, \& {the CANDELS
  Team}}]{galametz13}
{Galametz}, A., {Grazian}, A., {Fontana}, A., {et~al.} 2013, \apjs, 206, 10

\bibitem[{{Genel} {et~al.}(2018){Genel}, {Nelson}, {Pillepich}, {Springel},
  {Pakmor}, {Weinberger}, {Hernquist}, {Naiman}, {Vogelsberger}, {Marinacci},
  \& {Torrey}}]{Genel+18}
{Genel}, S., {Nelson}, D., {Pillepich}, A., {et~al.} 2018, \mnras, 474, 3976

\bibitem[{{George}(2017)}]{George17}
{George}, K. 2017, \aap, 598, A45

\bibitem[{{George} \& {Zingade}(2015)}]{George15}
{George}, K. \& {Zingade}, K. 2015, \aap, 583, A103

\bibitem[{{Girardi} {et~al.}(2005){Girardi}, {Groenewegen}, {Hatziminaoglou},
  \& {da Costa}}]{Girardi05}
{Girardi}, L., {Groenewegen}, M.~A.~T., {Hatziminaoglou}, E., \& {da Costa}, L.
  2005, \aap, 436, 895

\bibitem[{{Graham} {et~al.}(2006){Graham}, {Merritt}, {Moore}, {Diemand}, \&
  {Terzi{\'c}}}]{graham06}
{Graham}, A.~W., {Merritt}, D., {Moore}, B., {Diemand}, J., \& {Terzi{\'c}}, B.
  2006, \aj, 132, 2711

\bibitem[{{Grogin} {et~al.}(2011){Grogin}, {Kocevski}, {Faber}, {Ferguson},
  {Koekemoer}, {Riess}, {Acquaviva}, {Alexander}, {Almaini}, {Ashby}, {Barden},
  {Bell}, {Bournaud}, {Brown}, {Caputi}, {Casertano}, {Cassata}, {Castellano},
  {Challis}, {Chary}, {Cheung}, {Cirasuolo}, {Conselice}, {Roshan Cooray},
  {Croton}, {Daddi}, {Dahlen}, {Dav{\'e}}, {de Mello}, {Dekel}, {Dickinson},
  {Dolch}, {Donley}, {Dunlop}, {Dutton}, {Elbaz}, {Fazio}, {Filippenko},
  {Finkelstein}, {Fontana}, {Gardner}, {Garnavich}, {Gawiser}, {Giavalisco},
  {Grazian}, {Guo}, {Hathi}, {H{\"a}ussler}, {Hopkins}, {Huang}, {Huang},
  {Jha}, {Kartaltepe}, {Kirshner}, {Koo}, {Lai}, {Lee}, {Li}, {Lotz}, {Lucas},
  {Madau}, {McCarthy}, {McGrath}, {McIntosh}, {McLure}, {Mobasher},
  {Moustakas}, {Mozena}, {Nandra}, {Newman}, {Niemi}, {Noeske}, {Papovich},
  {Pentericci}, {Pope}, {Primack}, {Rajan}, {Ravindranath}, {Reddy}, {Renzini},
  {Rix}, {Robaina}, {Rodney}, {Rosario}, {Rosati}, {Salimbeni}, {Scarlata},
  {Siana}, {Simard}, {Smidt}, {Somerville}, {Spinrad}, {Straughn}, {Strolger},
  {Telford}, {Teplitz}, {Trump}, {van der Wel}, {Villforth}, {Wechsler},
  {Weiner}, {Wiklind}, {Wild}, {Wilson}, {Wuyts}, {Yan}, \& {Yun}}]{grogin11}
{Grogin}, N.~A., {Kocevski}, D.~D., {Faber}, S.~M., {et~al.} 2011, \apjs, 197,
  35

\bibitem[{{Grylls} {et~al.}(2019){Grylls}, {Shankar}, {Zanisi}, \&
  {Bernardi}}]{Grylls19}
{Grylls}, P.~J., {Shankar}, F., {Zanisi}, L., \& {Bernardi}, M. 2019, \mnras,
  483, 2506

\bibitem[{{Gu} {et~al.}(2020){Gu}, {Fang}, {Yuan}, \& {Lu}}]{Gu20}
{Gu}, Y., {Fang}, G., {Yuan}, Q., \& {Lu}, S. 2020, \pasp, 132, 054101

\bibitem[{{Gunn} \& {Gott}(1972)}]{1972ApJ...176....1G}
{Gunn}, J.~E. \& {Gott}, III, J.~R. 1972, \apj, 176, 1

\bibitem[{{Guo} {et~al.}(2011){Guo}, {White}, {Boylan-Kolchin}, {De Lucia},
  {Kauffmann}, {Lemson}, {Li}, {Springel}, \& {Weinmann}}]{Guo+11}
{Guo}, Q., {White}, S., {Boylan-Kolchin}, M., {et~al.} 2011, \mnras, 413, 101

\bibitem[{{Guo} {et~al.}(2013){Guo}, {Ferguson}, {Giavalisco}, {Barro},
  {Willner}, {Ashby}, {Dahlen}, {Donley}, {Faber}, {Fontana}, {Galametz},
  {Grazian}, {Huang}, {Kocevski}, {Koekemoer}, {Koo}, {McGrath}, {Peth},
  {Salvato}, {Wuyts}, {Castellano}, {Cooray}, {Dickinson}, {Dunlop}, {Fazio},
  {Gardner}, {Gawiser}, {Grogin}, {Hathi}, {Hsu}, {Lee}, {Lucas}, {Mobasher},
  {Nandra}, {Newman}, \& {van der Wel}}]{guo13}
{Guo}, Y., {Ferguson}, H.~C., {Giavalisco}, M., {et~al.} 2013, \apjs, 207, 24

\bibitem[{{Guo} {et~al.}(2009){Guo}, {McIntosh}, {Mo}, {Katz}, {Van Den Bosch},
  {Weinberg}, {Weinmann}, {Pasquali}, \& {Yang}}]{Guo09}
{Guo}, Y., {McIntosh}, D.~H., {Mo}, H.~J., {et~al.} 2009, \mnras, 398, 1129

\bibitem[{{Hamadouche} {et~al.}(2022){Hamadouche}, {Carnall}, {McLure},
  {Dunlop}, {McLeod}, {Cullen}, {Begley}, {Bolzonella}, {Buitrago},
  {Castellano}, {Cucciati}, {Fontana}, {Gargiulo}, {Moresco}, {Pozzetti}, \&
  {Zamorani}}]{Hamadouche22}
{Hamadouche}, M.~L., {Carnall}, A.~C., {McLure}, R.~J., {et~al.} 2022, \mnras,
  512, 1262

\bibitem[{{Hopkins} {et~al.}(2009{\natexlab{a}}){Hopkins}, {Bundy}, {Murray},
  {Quataert}, {Lauer}, \& {Ma}}]{Hopkins09}
{Hopkins}, P.~F., {Bundy}, K., {Murray}, N., {et~al.} 2009{\natexlab{a}},
  \mnras, 398, 898

\bibitem[{{Hopkins} {et~al.}(2009{\natexlab{b}}){Hopkins}, {Hernquist}, {Cox},
  {Keres}, \& {Wuyts}}]{2009ApJ...691.1424H}
{Hopkins}, P.~F., {Hernquist}, L., {Cox}, T.~J., {Keres}, D., \& {Wuyts}, S.
  2009{\natexlab{b}}, \apj, 691, 1424

\bibitem[{{Huang} {et~al.}(2018){Huang}, {Leauthaud}, {Greene}, {Bundy}, {Lin},
  {Tanaka}, {Mandelbaum}, {Miyazaki}, \& {Komiyama}}]{Huang18}
{Huang}, S., {Leauthaud}, A., {Greene}, J., {et~al.} 2018, \mnras, 480, 521

\bibitem[{{Huertas-Company} {et~al.}(2010){Huertas-Company}, {Aguerri},
  {Tresse}, {Bolzonella}, {Koekemoer}, \& {Maier}}]{Huertas10}
{Huertas-Company}, M., {Aguerri}, J.~A.~L., {Tresse}, L., {et~al.} 2010, \aap,
  515, A3

\bibitem[{{Huertas-Company} {et~al.}(2013{\natexlab{a}}){Huertas-Company},
  {Mei}, {Shankar}, {Delaye}, {Raichoor}, {Covone}, {Finoguenov}, {Kneib},
  {Le}, \& {Povic}}]{Huertas13a}
{Huertas-Company}, M., {Mei}, S., {Shankar}, F., {et~al.} 2013{\natexlab{a}},
  \mnras, 428, 1715

\bibitem[{{Huertas-Company} {et~al.}(2013{\natexlab{b}}){Huertas-Company},
  {Shankar}, {Mei}, {Bernardi}, {Aguerri}, {Meert}, \& {Vikram}}]{Huertas13b}
{Huertas-Company}, M., {Shankar}, F., {Mei}, S., {et~al.} 2013{\natexlab{b}},
  \apj, 779, 29

\bibitem[{{Jaff{\'e}} {et~al.}(2011){Jaff{\'e}}, {Arag{\'o}n-Salamanca}, {De
  Lucia}, {Jablonka}, {Rudnick}, {Saglia}, \& {Zaritsky}}]{Jaffe11}
{Jaff{\'e}}, Y.~L., {Arag{\'o}n-Salamanca}, A., {De Lucia}, G., {et~al.} 2011,
  \mnras, 410, 280

\bibitem[{{Kartaltepe} {et~al.}(2015){Kartaltepe}, {Mozena}, {Kocevski},
  {McIntosh}, {Lotz}, {Bell}, {Faber}, {Ferguson}, {Koo}, {Bassett}, {Bernyk},
  {Blancato}, {Bournaud}, {Cassata}, {Castellano}, {Cheung}, {Conselice},
  {Croton}, {Dahlen}, {de Mello}, {DeGroot}, {Donley}, {Guedes}, {Grogin},
  {Hathi}, {Hilton}, {Hollon}, {Koekemoer}, {Liu}, {Lucas}, {Martig},
  {McGrath}, {McPartland}, {Mobasher}, {Morlock}, {O'Leary}, {Peth}, {Pforr},
  {Pillepich}, {Rosario}, {Soto}, {Straughn}, {Telford}, {Sunnquist}, {Trump},
  {Weiner}, {Wuyts}, {Inami}, {Kassin}, {Lani}, {Poole}, \&
  {Rizer}}]{Kartaltepe15}
{Kartaltepe}, J.~S., {Mozena}, M., {Kocevski}, D., {et~al.} 2015, \apjs, 221,
  11

\bibitem[{{Kauffmann} {et~al.}(2003){Kauffmann}, {Heckman}, {White}, {Charlot},
  {Tremonti}, {Brinchmann}, {Bruzual}, {Peng}, {Seibert}, {Bernardi},
  {Blanton}, {Brinkmann}, {Castander}, {Cs{\'a}bai}, {Fukugita}, {Ivezic},
  {Munn}, {Nichol}, {Padmanabhan}, {Thakar}, {Weinberg}, \&
  {York}}]{Kaufmann03}
{Kauffmann}, G., {Heckman}, T.~M., {White}, S. D.~M., {et~al.} 2003, \mnras,
  341, 33

\bibitem[{{Kelkar} {et~al.}(2015){Kelkar}, {Arag{\'o}n-Salamanca}, {Gray},
  {Maltby}, {Vulcani}, {De Lucia}, {Poggianti}, \& {Zaritsky}}]{Kelkar15}
{Kelkar}, K., {Arag{\'o}n-Salamanca}, A., {Gray}, M.~E., {et~al.} 2015, \mnras,
  450, 1246

\bibitem[{{Koekemoer} {et~al.}(2011){Koekemoer}, {Faber}, {Ferguson}, {Grogin},
  {Kocevski}, {Koo}, {Lai}, {Lotz}, {Lucas}, {McGrath}, {Ogaz}, {Rajan},
  {Riess}, {Rodney}, {Strolger}, {Casertano}, {Castellano}, {Dahlen},
  {Dickinson}, {Dolch}, {Fontana}, {Giavalisco}, {Grazian}, {Guo}, {Hathi},
  {Huang}, {van der Wel}, {Yan}, {Acquaviva}, {Alexander}, {Almaini}, {Ashby},
  {Barden}, {Bell}, {Bournaud}, {Brown}, {Caputi}, {Cassata}, {Challis},
  {Chary}, {Cheung}, {Cirasuolo}, {Conselice}, {Roshan Cooray}, {Croton},
  {Daddi}, {Dav{\'e}}, {de Mello}, {de Ravel}, {Dekel}, {Donley}, {Dunlop},
  {Dutton}, {Elbaz}, {Fazio}, {Filippenko}, {Finkelstein}, {Frazer}, {Gardner},
  {Garnavich}, {Gawiser}, {Gruetzbauch}, {Hartley}, {H{\"a}ussler},
  {Herrington}, {Hopkins}, {Huang}, {Jha}, {Johnson}, {Kartaltepe},
  {Khostovan}, {Kirshner}, {Lani}, {Lee}, {Li}, {Madau}, {McCarthy},
  {McIntosh}, {McLure}, {McPartland}, {Mobasher}, {Moreira}, {Mortlock},
  {Moustakas}, {Mozena}, {Nandra}, {Newman}, {Nielsen}, {Niemi}, {Noeske},
  {Papovich}, {Pentericci}, {Pope}, {Primack}, {Ravindranath}, {Reddy},
  {Renzini}, {Rix}, {Robaina}, {Rosario}, {Rosati}, {Salimbeni}, {Scarlata},
  {Siana}, {Simard}, {Smidt}, {Snyder}, {Somerville}, {Spinrad}, {Straughn},
  {Telford}, {Teplitz}, {Trump}, {Vargas}, {Villforth}, {Wagner}, {Wandro},
  {Wechsler}, {Weiner}, {Wiklind}, {Wild}, {Wilson}, {Wuyts}, \&
  {Yun}}]{koekemoer11}
{Koekemoer}, A.~M., {Faber}, S.~M., {Ferguson}, H.~C., {et~al.} 2011, \apjs,
  197, 36

\bibitem[{{Kravtsov}(2013)}]{Kravtsov13}
{Kravtsov}, A.~V. 2013, \apjl, 764, L31

\bibitem[{{Kuchner} {et~al.}(2017){Kuchner}, {Ziegler}, {Verdugo}, {Bamford},
  \& {H{\"a}u{\ss}ler}}]{Kuchner17}
{Kuchner}, U., {Ziegler}, B., {Verdugo}, M., {Bamford}, S., \&
  {H{\"a}u{\ss}ler}, B. 2017, \aap, 604, A54

\bibitem[{{K{\"u}mmel} {et~al.}(2009){K{\"u}mmel}, {Walsh}, {Pirzkal},
  {Kuntschner}, \& {Pasquali}}]{Kummel09}
{K{\"u}mmel}, M., {Walsh}, J.~R., {Pirzkal}, N., {Kuntschner}, H., \&
  {Pasquali}, A. 2009, \pasp, 121, 59

\bibitem[{{Labb{\'e}} {et~al.}(2005){Labb{\'e}}, {Huang}, {Franx}, {Rudnick},
  {Barmby}, {Daddi}, {van Dokkum}, {Fazio}, {F{\"o}rster Schreiber},
  {Moorwood}, {Rix}, {R{\"o}ttgering}, {Trujillo}, \& {van der Werf}}]{Labbe05}
{Labb{\'e}}, I., {Huang}, J., {Franx}, M., {et~al.} 2005, \apjl, 624, L81

\bibitem[{{Lacy} {et~al.}(2005){Lacy}, {Wilson}, {Masci}, {Storrie-Lombardi},
  {Appleton}, {Armus}, {Chapman}, {Choi}, {Fadda}, {Fang}, {Frayer},
  {Heinrichsen}, {Helou}, {Im}, {Laine}, {Marleau}, {Shupe}, {Soifer},
  {Squires}, {Surace}, {Teplitz}, \& {Yan}}]{Lacy05}
{Lacy}, M., {Wilson}, G., {Masci}, F., {et~al.} 2005, \apjs, 161, 41

\bibitem[{{Lange} {et~al.}(2015){Lange}, {Driver}, {Robotham}, {Kelvin},
  {Graham}, {Alpaslan}, {Andrews}, {Baldry}, {Bamford}, {Bland-Hawthorn},
  {Brough}, {Cluver}, {Conselice}, {Davies}, {Haeussler}, {Konstantopoulos},
  {Loveday}, {Moffett}, {Norberg}, {Phillipps}, {Taylor},
  {L{\'o}pez-S{\'a}nchez}, \& {Wilkins}}]{Lange15}
{Lange}, R., {Driver}, S.~P., {Robotham}, A. S.~G., {et~al.} 2015, \mnras, 447,
  2603

\bibitem[{{Lapi} {et~al.}(2018){Lapi}, {Pantoni}, {Zanisi}, {Shi}, {Mancuso},
  {Massardi}, {Shankar}, {Bressan}, \& {Danese}}]{Lapi18}
{Lapi}, A., {Pantoni}, L., {Zanisi}, L., {et~al.} 2018, \apj, 857, 22

\bibitem[{{Lee} {et~al.}(2006){Lee}, {Lee}, \& {Hwang}}]{Lee06}
{Lee}, J.~H., {Lee}, M.~G., \& {Hwang}, H.~S. 2006, \apj, 650, 148

\bibitem[{{Li} {et~al.}(2018){Li}, {Mao}, {Cappellari}, {Ge}, {Long}, {Li},
  {Mo}, {Li}, {Zheng}, {Bundy}, {Thomas}, {Brownstein}, {Roman Lopes}, {Law},
  \& {Drory}}]{li18}
{Li}, H., {Mao}, S., {Cappellari}, M., {et~al.} 2018, \mnras, 476, 1765

\bibitem[{{Lilly} \& {Carollo}(2016)}]{Lilly16}
{Lilly}, S.~J. \& {Carollo}, C.~M. 2016, \apj, 833, 1

\bibitem[{{Lu} {et~al.}(2019){Lu}, {Gu}, {Fang}, \& {Yuan}}]{Lu19}
{Lu}, S.-Y., {Gu}, Y.-Z., {Fang}, G.-W., \& {Yuan}, Q.-R. 2019, Research in
  Astronomy and Astrophysics, 19, 150

\bibitem[{{Lustig} {et~al.}(2021){Lustig}, {Strazzullo}, {D'Eugenio}, {Daddi},
  {Pannella}, {Renzini}, {Cimatti}, {Gobat}, {Jin}, {Mohr}, \&
  {Onodera}}]{Lustig21}
{Lustig}, P., {Strazzullo}, V., {D'Eugenio}, C., {et~al.} 2021, \mnras, 501,
  2659

\bibitem[{{Makovoz} \& {Khan}(2005)}]{Makovoz05}
{Makovoz}, D. \& {Khan}, I. 2005, in Astronomical Society of the Pacific
  Conference Series, Vol. 347, Astronomical Data Analysis Software and Systems
  XIV, ed. P.~{Shopbell}, M.~{Britton}, \& R.~{Ebert}, 81

\bibitem[{{Maltby} {et~al.}(2018){Maltby}, {Almaini}, {Wild}, {Hatch},
  {Hartley}, {Simpson}, {Rowlands}, \& {Socolovsky}}]{Maltby18}
{Maltby}, D.~T., {Almaini}, O., {Wild}, V., {et~al.} 2018, \mnras, 480, 381

\bibitem[{{Maltby} {et~al.}(2010){Maltby}, {Arag{\'o}n-Salamanca}, {Gray},
  {Barden}, {H{\"a}u{\ss}ler}, {Wolf}, {Peng}, {Jahnke}, {McIntosh},
  {B{\"o}hm}, \& {van Kampen}}]{Maltby10}
{Maltby}, D.~T., {Arag{\'o}n-Salamanca}, A., {Gray}, M.~E., {et~al.} 2010,
  \mnras, 402, 282

\bibitem[{{Mansheim} {et~al.}(2017){Mansheim}, {Lemaux}, {Dawson}, {Lubin},
  {Wittman}, \& {Schmidt}}]{Mansheim17}
{Mansheim}, A.~S., {Lemaux}, B.~C., {Dawson}, W.~A., {et~al.} 2017, \apj, 834,
  205

\bibitem[{{Marsan} {et~al.}(2019){Marsan}, {Marchesini}, {Muzzin}, {Brammer},
  {Bezanson}, {Franx}, {Labb{\'e}}, {Lundgren}, {Rudnick}, {Stefanon}, {van
  Dokkum}, {Wake}, \& {Whitaker}}]{Marsan19}
{Marsan}, Z.~C., {Marchesini}, D., {Muzzin}, A., {et~al.} 2019, \apj, 871, 201

\bibitem[{{Matharu} {et~al.}(2020){Matharu}, {Muzzin}, {Brammer}, {van der
  Burg}, {Auger}, {Hewett}, {Chan}, {Demarco}, {van Dokkum}, {Marchesini},
  {Nelson}, {Noble}, \& {Wilson}}]{Matharu20}
{Matharu}, J., {Muzzin}, A., {Brammer}, G.~B., {et~al.} 2020, \mnras, 493, 6011

\bibitem[{{Matharu} {et~al.}(2019){Matharu}, {Muzzin}, {Brammer}, {van der
  Burg}, {Auger}, {Hewett}, {van der Wel}, {van Dokkum}, {Balogh}, {Chan},
  {Demarco}, {Marchesini}, {Nelson}, {Noble}, {Wilson}, \& {Yee}}]{Matharu19}
{Matharu}, J., {Muzzin}, A., {Brammer}, G.~B., {et~al.} 2019, \mnras, 484, 595

\bibitem[{{Matteuzzi} {et~al.}(2022){Matteuzzi}, {Marinacci}, {Nipoti}, \&
  {Andreon}}]{Matteuzzi22}
{Matteuzzi}, M., {Marinacci}, F., {Nipoti}, C., \& {Andreon}, S. 2022, \mnras,
  513, 3893

\bibitem[{{McIntosh} {et~al.}(2005){McIntosh}, {Bell}, {Rix}, {Wolf},
  {Heymans}, {Peng}, {Somerville}, {Barden}, {Beckwith}, {Borch}, {Caldwell},
  {H{\"a}u{\ss}ler}, {Jahnke}, {Jogee}, {Meisenheimer}, {S{\'a}nchez}, \&
  {Wisotzki}}]{McIntosh05}
{McIntosh}, D.~H., {Bell}, E.~F., {Rix}, H.-W., {et~al.} 2005, \apj, 632, 191

\bibitem[{{Mei} {et~al.}(2006){Mei}, {Blakeslee}, {Stanford}, {Holden},
  {Rosati}, {Strazzullo}, {Homeier}, {Postman}, {Franx}, {Rettura}, {Ford},
  {Illingworth}, {Ettori}, {Bouwens}, {Demarco}, {Martel}, {Clampin}, {Hartig},
  {Eisenhardt}, {Ardila}, {Bartko}, {Ben{\'\i}tez}, {Bradley}, {Broadhurst},
  {Brown}, {Burrows}, {Cheng}, {Cross}, {Feldman}, {Golimowski}, {Goto},
  {Gronwall}, {Infante}, {Kimble}, {Krist}, {Lesser}, {Menanteau}, {Meurer},
  {Miley}, {Motta}, {Sirianni}, {Sparks}, {Tran}, {Tsvetanov}, {White}, \&
  {Zheng}}]{Mei06}
{Mei}, S., {Blakeslee}, J.~P., {Stanford}, S.~A., {et~al.} 2006, \apj, 639, 81

\bibitem[{{Mei} {et~al.}(2022){Mei}, {Hatch}, {Amodeo}, {Afanasiev}, {De
  Breuck}, {Stern}, {Cooke}, {Gonzalez}, {Noirot}, {Rettura}, {Seymour},
  {Stanford}, {Vernet}, \& {Wylezalek}}]{Mei22}
{Mei}, S., {Hatch}, N.~A., {Amodeo}, S., {et~al.} 2022, arXiv e-prints,
  arXiv:2209.02078

\bibitem[{{Mei} {et~al.}(2015){Mei}, {Scarlata}, {Pentericci}, {Newman},
  {Weiner}, {Ashby}, {Castellano}, {Conselice}, {Finkelstein}, {Galametz},
  {Grogin}, {Koekemoer}, {Huertas-Company}, {Lani}, {Lucas}, {Papovich},
  {Rafelski}, \& {Teplitz}}]{Mei15}
{Mei}, S., {Scarlata}, C., {Pentericci}, L., {et~al.} 2015, \apj, 804, 117

\bibitem[{{Merlin} {et~al.}(2016){Merlin}, {Bourne}, {Castellano}, {Ferguson},
  {Wang}, {Derriere}, {Dunlop}, {Elbaz}, \& {Fontana}}]{merlin+16}
{Merlin}, E., {Bourne}, N., {Castellano}, M., {et~al.} 2016, \aap, 595, A97

\bibitem[{{Merlin} {et~al.}(2015){Merlin}, {Fontana}, {Ferguson}, {Dunlop},
  {Elbaz}, {Bourne}, {Bruce}, {Buitrago}, {Castellano}, {Schreiber}, {Grazian},
  {McLure}, {Okumura}, {Shu}, {Wang}, {Amor{\'{\i}}n}, {Boutsia}, {Cappelluti},
  {Comastri}, {Derriere}, {Faber}, \& {Santini}}]{merlin+15}
{Merlin}, E., {Fontana}, A., {Ferguson}, H.~C., {et~al.} 2015, \aap, 582, A15

\bibitem[{{Miller} {et~al.}(2022){Miller}, {van Dokkum}, \& {Mowla}}]{Miller22}
{Miller}, T.~B., {van Dokkum}, P., \& {Mowla}, L. 2022, arXiv e-prints,
  arXiv:2207.05895

\bibitem[{{Miller} {et~al.}(2019){Miller}, {van Dokkum}, {Mowla}, \& {van der
  Wel}}]{Miller19}
{Miller}, T.~B., {van Dokkum}, P., {Mowla}, L., \& {van der Wel}, A. 2019,
  \apjl, 872, L14

\bibitem[{{Moore} {et~al.}(1996){Moore}, {Katz}, {Lake}, {Dressler}, \&
  {Oemler}}]{Moore96}
{Moore}, B., {Katz}, N., {Lake}, G., {Dressler}, A., \& {Oemler}, A. 1996,
  \nat, 379, 613

\bibitem[{{Mosleh} {et~al.}(2018){Mosleh}, {Tavasoli}, \&
  {Tacchella}}]{Mosleh18}
{Mosleh}, M., {Tavasoli}, S., \& {Tacchella}, S. 2018, \apj, 861, 101

\bibitem[{{Mosleh} {et~al.}(2011){Mosleh}, {Williams}, {Franx}, \&
  {Kriek}}]{Mosleh11}
{Mosleh}, M., {Williams}, R.~J., {Franx}, M., \& {Kriek}, M. 2011, \apj, 727, 5

\bibitem[{{Moster} {et~al.}(2018){Moster}, {Naab}, \& {White}}]{Moster18}
{Moster}, B.~P., {Naab}, T., \& {White}, S. D.~M. 2018, \mnras, 477, 1822

\bibitem[{{Mowla} {et~al.}(2019{\natexlab{a}}){Mowla}, {van der Wel}, {van
  Dokkum}, \& {Miller}}]{Mowla19a}
{Mowla}, L., {van der Wel}, A., {van Dokkum}, P., \& {Miller}, T.~B.
  2019{\natexlab{a}}, \apjl, 872, L13

\bibitem[{{Mowla} {et~al.}(2019{\natexlab{b}}){Mowla}, {van Dokkum}, {Brammer},
  {Momcheva}, {van der Wel}, {Whitaker}, {Nelson}, {Bezanson}, {Muzzin},
  {Franx}, {MacKenty}, {Leja}, {Kriek}, \& {Marchesini}}]{Mowla19}
{Mowla}, L.~A., {van Dokkum}, P., {Brammer}, G.~B., {et~al.}
  2019{\natexlab{b}}, \apj, 880, 57

\bibitem[{{Muldrew} {et~al.}(2015){Muldrew}, {Hatch}, \& {Cooke}}]{Muldrew15}
{Muldrew}, S.~I., {Hatch}, N.~A., \& {Cooke}, E.~A. 2015, \mnras, 452, 2528

\bibitem[{{Naab} {et~al.}(2009){Naab}, {Johansson}, \& {Ostriker}}]{Naab09}
{Naab}, T., {Johansson}, P.~H., \& {Ostriker}, J.~P. 2009, \apjl, 699, L178

\bibitem[{{Nair} {et~al.}(2011){Nair}, {van den Bergh}, \& {Abraham}}]{Nair11}
{Nair}, P., {van den Bergh}, S., \& {Abraham}, R.~G. 2011, \apjl, 734, L31

\bibitem[{{Nedkova} {et~al.}(2021){Nedkova}, {H{\"a}u{\ss}ler}, {Marchesini},
  {Dimauro}, {Brammer}, {Eigenthaler}, {Feinstein}, {Ferguson},
  {Huertas-Company}, {Johnston}, {Kado-Fong}, {Kartaltepe}, {Labb{\'e}},
  {Lange-Vagle}, {Martis}, {McGrath}, {Muzzin}, {Oesch}, {Ordenes-Brice{\~n}o},
  {Puzia}, {Shipley}, {Simmons}, {Skelton}, {Stefanon}, {van der Wel}, \&
  {Whitaker}}]{Nedkova21}
{Nedkova}, K.~V., {H{\"a}u{\ss}ler}, B., {Marchesini}, D., {et~al.} 2021,
  \mnras, 506, 928

\bibitem[{{Newman} {et~al.}(2014){Newman}, {Ellis}, {Andreon}, {Treu},
  {Raichoor}, \& {Trinchieri}}]{Newman14}
{Newman}, A.~B., {Ellis}, R.~S., {Andreon}, S., {et~al.} 2014, \apj, 788, 51

\bibitem[{{Newman} {et~al.}(2012){Newman}, {Ellis}, {Bundy}, \&
  {Treu}}]{Newman12}
{Newman}, A.~B., {Ellis}, R.~S., {Bundy}, K., \& {Treu}, T. 2012, \apj, 746,
  162

\bibitem[{{Noirot} {et~al.}(2018){Noirot}, {Stern}, {Mei}, {Wylezalek},
  {Cooke}, {De Breuck}, {Galametz}, {Hatch}, {Vernet}, {Brodwin}, {Eisenhardt},
  {Gonzalez}, {Jarvis}, {Rettura}, {Seymour}, \& {Stanford}}]{Noirot18}
{Noirot}, G., {Stern}, D., {Mei}, S., {et~al.} 2018, \apj, 859, 38

\bibitem[{{Noirot} {et~al.}(2016){Noirot}, {Vernet}, {De Breuck}, {Wylezalek},
  {Galametz}, {Stern}, {Mei}, {Brodwin}, {Cooke}, {Gonzalez}, {Hatch},
  {Rettura}, \& {Stanford}}]{Noirot16}
{Noirot}, G., {Vernet}, J., {De Breuck}, C., {et~al.} 2016, \apj, 830, 90

\bibitem[{{Noordeh} {et~al.}(2021){Noordeh}, {Canning}, {Willis}, {Allen},
  {Mantz}, {Stanford}, \& {Brammer}}]{Noordeh21}
{Noordeh}, E., {Canning}, R.~E.~A., {Willis}, J.~P., {et~al.} 2021, \mnras,
  507, 5272

\bibitem[{{Papovich} {et~al.}(2012){Papovich}, {Bassett}, {Lotz}, {van der
  Wel}, {Tran}, {Finkelstein}, {Bell}, {Conselice}, {Dekel}, {Dunlop}, {Guo},
  {Faber}, {Farrah}, {Ferguson}, {Finkelstein}, {H{\"a}ussler}, {Kocevski},
  {Koekemoer}, {Koo}, {McGrath}, {McLure}, {McIntosh}, {Momcheva}, {Newman},
  {Rudnick}, {Weiner}, {Willmer}, \& {Wuyts}}]{Papovich12}
{Papovich}, C., {Bassett}, R., {Lotz}, J.~M., {et~al.} 2012, \apj, 750, 93

\bibitem[{{Patel} {et~al.}(2017){Patel}, {Hong}, {Quadri}, {Holden}, \&
  {Williams}}]{Patel17}
{Patel}, S.~G., {Hong}, Y.~X., {Quadri}, R.~F., {Holden}, B.~P., \& {Williams},
  R.~J. 2017, \apj, 839, 127

\bibitem[{{Peng} {et~al.}(2002){Peng}, {Ho}, {Impey}, \& {Rix}}]{peng02}
{Peng}, C.~Y., {Ho}, L.~C., {Impey}, C.~D., \& {Rix}, H.-W. 2002, \aj, 124, 266

\bibitem[{{Poggianti} {et~al.}(2013){Poggianti}, {Calvi}, {Bindoni},
  {D'Onofrio}, {Moretti}, {Valentinuzzi}, {Fasano}, {Fritz}, {De Lucia},
  {Vulcani}, {Bettoni}, {Gullieuszik}, \& {Omizzolo}}]{Poggianti13a}
{Poggianti}, B.~M., {Calvi}, R., {Bindoni}, D., {et~al.} 2013, \apj, 762, 77

\bibitem[{{Postman} {et~al.}(2005){Postman}, {Franx}, {Cross}, {Holden},
  {Ford}, {Illingworth}, {Goto}, {Demarco}, {Rosati}, {Blakeslee}, {Tran},
  {Ben{\'\i}tez}, {Clampin}, {Hartig}, {Homeier}, {Ardila}, {Bartko},
  {Bouwens}, {Bradley}, {Broadhurst}, {Brown}, {Burrows}, {Cheng}, {Feldman},
  {Golimowski}, {Gronwall}, {Infante}, {Kimble}, {Krist}, {Lesser}, {Martel},
  {Mei}, {Menanteau}, {Meurer}, {Miley}, {Motta}, {Sirianni}, {Sparks}, {Tran},
  {Tsvetanov}, {White}, \& {Zheng}}]{Postman05}
{Postman}, M., {Franx}, M., {Cross}, N.~J.~G., {et~al.} 2005, \apj, 623, 721

\bibitem[{{Raichoor} {et~al.}(2012){Raichoor}, {Mei}, {Stanford}, {Holden},
  {Nakata}, {Rosati}, {Shankar}, {Tanaka}, {Ford}, {Huertas-Company},
  {Illingworth}, {Kodama}, {Postman}, {Rettura}, {Blakeslee}, {Demarco}, {Jee},
  \& {White}}]{Raichoor12}
{Raichoor}, A., {Mei}, S., {Stanford}, S.~A., {et~al.} 2012, \apj, 745, 130

\bibitem[{{Rettura} {et~al.}(2010){Rettura}, {Rosati}, {Nonino}, {Fosbury},
  {Gobat}, {Menci}, {Strazzullo}, {Mei}, {Demarco}, \& {Ford}}]{Rettura10}
{Rettura}, A., {Rosati}, P., {Nonino}, M., {et~al.} 2010, \apj, 709, 512

\bibitem[{{Rieke} {et~al.}(2004){Rieke}, {Young}, {Engelbracht}, {Kelly},
  {Low}, {Haller}, {Beeman}, {Gordon}, {Stansberry}, {Misselt}, {Cadien},
  {Morrison}, {Rivlis}, {Latter}, {Noriega-Crespo}, {Padgett}, {Stapelfeldt},
  {Hines}, {Egami}, {Muzerolle}, {Alonso-Herrero}, {Blaylock}, {Dole}, {Hinz},
  {Le Floc'h}, {Papovich}, {P{\'e}rez-Gonz{\'a}lez}, {Smith}, {Su}, {Bennett},
  {Frayer}, {Henderson}, {Lu}, {Masci}, {Pesenson}, {Rebull}, {Rho}, {Keene},
  {Stolovy}, {Wachter}, {Wheaton}, {Werner}, \& {Richards}}]{Rieke04}
{Rieke}, G.~H., {Young}, E.~T., {Engelbracht}, C.~W., {et~al.} 2004, \apjs,
  154, 25

\bibitem[{{Rohr} {et~al.}(2022){Rohr}, {Feldmann}, {Bullock},
  {{\c{C}}atmabacak}, {Boylan-Kolchin}, {Faucher-Gigu{\`e}re}, {Kere{\v{s}}},
  {Liang}, {Moreno}, \& {Wetzel}}]{Rohr22}
{Rohr}, E., {Feldmann}, R., {Bullock}, J.~S., {et~al.} 2022, \mnras, 510, 3967

\bibitem[{{Saglia} {et~al.}(2010){Saglia}, {S{\'a}nchez-Bl{\'a}zquez},
  {Bender}, {Simard}, {Desai}, {Arag{\'o}n-Salamanca}, {Milvang-Jensen},
  {Halliday}, {Jablonka}, {Noll}, {Poggianti}, {Clowe}, {De Lucia},
  {Pell{\'o}}, {Rudnick}, {Valentinuzzi}, {White}, \& {Zaritsky}}]{Saglia10}
{Saglia}, R.~P., {S{\'a}nchez-Bl{\'a}zquez}, P., {Bender}, R., {et~al.} 2010,
  \aap, 524, A6

\bibitem[{{Santini} {et~al.}(2015){Santini}, {Ferguson}, {Fontana}, {Mobasher},
  {Barro}, {Castellano}, {Finkelstein}, {Grazian}, {Hsu}, {Lee}, {Lee},
  {Pforr}, {Salvato}, {Wiklind}, {Wuyts}, {Almaini}, {Cooper}, {Galametz},
  {Weiner}, {Amorin}, {Boutsia}, {Conselice}, {Dahlen}, {Dickinson},
  {Giavalisco}, {Grogin}, {Guo}, {Hathi}, {Kocevski}, {Koekemoer},
  {Kurczynski}, {Merlin}, {Mortlock}, {Newman}, {Paris}, {Pentericci},
  {Simons}, \& {Willner}}]{Santini15}
{Santini}, P., {Ferguson}, H.~C., {Fontana}, A., {et~al.} 2015, \apj, 801, 97

\bibitem[{{Saracco} {et~al.}(2017){Saracco}, {Gargiulo}, {Ciocca}, \&
  {Marchesini}}]{Saracco17}
{Saracco}, P., {Gargiulo}, A., {Ciocca}, F., \& {Marchesini}, D. 2017, \aap,
  597, A122

\bibitem[{{Schirmer}(2013)}]{Schirmer13}
{Schirmer}, M. 2013, \apjs, 209, 21

\bibitem[{{Sersic}(1968)}]{sersic}
{Sersic}, J.~L. 1968, {Atlas de Galaxias Australes}

\bibitem[{{Shankar} \& {Bernardi}(2009)}]{ShankarBernardi09}
{Shankar}, F. \& {Bernardi}, M. 2009, \mnras, 396, L76

\bibitem[{{Shankar} {et~al.}(2015){Shankar}, {Buchan}, {Rettura}, {Bouillot},
  {Moreno}, {Licitra}, {Bernardi}, {Huertas-Company}, {Mei}, {Ascaso}, {Sheth},
  {Delaye}, \& {Raichoor}}]{Shankar15}
{Shankar}, F., {Buchan}, S., {Rettura}, A., {et~al.} 2015, \apj, 802, 73

\bibitem[{{Shankar} {et~al.}(2013){Shankar}, {Marulli}, {Bernardi}, {Mei},
  {Meert}, \& {Vikram}}]{Shankar13}
{Shankar}, F., {Marulli}, F., {Bernardi}, M., {et~al.} 2013, \mnras, 428, 109

\bibitem[{{Shankar} {et~al.}(2014){Shankar}, {Mei}, {Huertas-Company},
  {Moreno}, {Fontanot}, {Monaco}, {Bernardi}, {Cattaneo}, {Sheth}, {Licitra},
  {Delaye}, \& {Raichoor}}]{Shankar14a}
{Shankar}, F., {Mei}, S., {Huertas-Company}, M., {et~al.} 2014, \mnras, 439,
  3189

\bibitem[{{Sheen} {et~al.}(2016){Sheen}, {Yi}, {Ree}, {Jaff{\'e}}, {Demarco},
  \& {Treister}}]{Sheen16}
{Sheen}, Y.-K., {Yi}, S.~K., {Ree}, C.~H., {et~al.} 2016, \apj, 827, 32

\bibitem[{{Sil'chenko} {et~al.}(2018){Sil'chenko}, {Kniazev}, \&
  {Chudakova}}]{Silchenko18}
{Sil'chenko}, O.~K., {Kniazev}, A.~Y., \& {Chudakova}, E.~M. 2018, \aj, 156,
  118

\bibitem[{{Socolovsky} {et~al.}(2019){Socolovsky}, {Maltby}, {Hatch},
  {Almaini}, {Wild}, {Hartley}, {Simpson}, \& {Rowlands}}]{Socolovsky19}
{Socolovsky}, M., {Maltby}, D.~T., {Hatch}, N.~A., {et~al.} 2019, \mnras, 482,
  1640

\bibitem[{{Somerville} {et~al.}(2018){Somerville}, {Behroozi}, {Pandya},
  {Dekel}, {Faber}, {Fontana}, {Koekemoer}, {Koo}, {P{\'e}rez-Gonz{\'a}lez},
  {Primack}, {Santini}, {Taylor}, \& {van der Wel}}]{Somerville18}
{Somerville}, R.~S., {Behroozi}, P., {Pandya}, V., {et~al.} 2018, \mnras, 473,
  2714

\bibitem[{{Strazzullo} {et~al.}(2013){Strazzullo}, {Gobat}, {Daddi}, {Onodera},
  {Carollo}, {Dickinson}, {Renzini}, {Arimoto}, {Cimatti}, {Finoguenov}, \&
  {Chary}}]{Strazzullo13}
{Strazzullo}, V., {Gobat}, R., {Daddi}, E., {et~al.} 2013, \apj, 772, 118

\bibitem[{{Strazzullo} {et~al.}(2022){Strazzullo}, {Pannella}, {Mohr}, {Saro},
  {Ashby}, {Bayliss}, {Canning}, {Floyd}, {Gonzalez}, {Khullar}, {Kim},
  {McDonald}, {Reichardt}, {Sharon}, \& {Somboonpanyakul}}]{Strazzullo22}
{Strazzullo}, V., {Pannella}, M., {Mohr}, J.~J., {et~al.} 2022, arXiv e-prints,
  arXiv:2212.06853

\bibitem[{{Stringer} {et~al.}(2014){Stringer}, {Shankar}, {Novak},
  {Huertas-Company}, {Combes}, \& {Moster}}]{Stringer14}
{Stringer}, M.~J., {Shankar}, F., {Novak}, G.~S., {et~al.} 2014, \mnras, 441,
  1570

\bibitem[{{Suess} {et~al.}(2019){Suess}, {Kriek}, {Price}, \&
  {Barro}}]{Suess19}
{Suess}, K.~A., {Kriek}, M., {Price}, S.~H., \& {Barro}, G. 2019, \apj, 877,
  103

\bibitem[{{Szomoru} {et~al.}(2012){Szomoru}, {Franx}, \& {van
  Dokkum}}]{Szomoru12}
{Szomoru}, D., {Franx}, M., \& {van Dokkum}, P.~G. 2012, \apj, 749, 121

\bibitem[{{Tadaki} {et~al.}(2020){Tadaki}, {Belli}, {Burkert}, {Dekel},
  {F{\"o}rster Schreiber}, {Genzel}, {Hayashi}, {Herrera-Camus}, {Kodama},
  {Kohno}, {Koyama}, {Lee}, {Lutz}, {Mowla}, {Nelson}, {Renzini}, {Suzuki},
  {Tacconi}, {{\"U}bler}, {Wisnioski}, \& {Wuyts}}]{Tadaki20}
{Tadaki}, K.-i., {Belli}, S., {Burkert}, A., {et~al.} 2020, \apj, 901, 74

\bibitem[{{Thomas} {et~al.}(2005){Thomas}, {Maraston}, {Bender}, \& {Mendes de
  Oliveira}}]{Thomas05}
{Thomas}, D., {Maraston}, C., {Bender}, R., \& {Mendes de Oliveira}, C. 2005,
  \apj, 621, 673

\bibitem[{{Trujillo} {et~al.}(2011){Trujillo}, {Ferreras}, \& {de La
  Rosa}}]{Trujillo11}
{Trujillo}, I., {Ferreras}, I., \& {de La Rosa}, I.~G. 2011, \mnras, 415, 3903

\bibitem[{{Trujillo} {et~al.}(2006){Trujillo}, {F{\"o}rster Schreiber},
  {Rudnick}, {Barden}, {Franx}, {Rix}, {Caldwell}, {McIntosh}, {Toft},
  {H{\"a}ussler}, {Zirm}, {van Dokkum}, {Labb{\'e}}, {Moorwood},
  {R{\"o}ttgering}, {van der Wel}, {van der Werf}, \& {van
  Starkenburg}}]{Trujillo06}
{Trujillo}, I., {F{\"o}rster Schreiber}, N.~M., {Rudnick}, G., {et~al.} 2006,
  \apj, 650, 18

\bibitem[{{Trujillo} {et~al.}(2004){Trujillo}, {Rudnick}, {Rix}, {Labb{\'e}},
  {Franx}, {Daddi}, {van Dokkum}, {F{\"o}rster Schreiber}, {Kuijken},
  {Moorwood}, {R{\"o}ttgering}, {van der Wel}, {van der Werf}, \& {van
  Starkenburg}}]{Trujillo04}
{Trujillo}, I., {Rudnick}, G., {Rix}, H.-W., {et~al.} 2004, \apj, 604, 521

\bibitem[{{van den Bosch} {et~al.}(2008){van den Bosch}, {Aquino}, {Yang},
  {Mo}, {Pasquali}, {McIntosh}, {Weinmann}, \& {Kang}}]{vandenBosch08}
{van den Bosch}, F.~C., {Aquino}, D., {Yang}, X., {et~al.} 2008, \mnras, 387,
  79

\bibitem[{{van der Wel} {et~al.}(2012){van der Wel}, {Bell}, {H{\"a}ussler},
  {McGrath}, {Chang}, {Guo}, {McIntosh}, {Rix}, {Barden}, {Cheung}, {Faber},
  {Ferguson}, {Galametz}, {Grogin}, {Hartley}, {Kartaltepe}, {Kocevski},
  {Koekemoer}, {Lotz}, {Mozena}, {Peth}, \& {Peng}}]{vanderWel12}
{van der Wel}, A., {Bell}, E.~F., {H{\"a}ussler}, B., {et~al.} 2012, \apjs,
  203, 24

\bibitem[{{van der Wel} {et~al.}(2014){van der Wel}, {Franx}, {van Dokkum},
  {Skelton}, {Momcheva}, {Whitaker}, {Brammer}, {Bell}, {Rix}, {Wuyts},
  {Ferguson}, {Holden}, {Barro}, {Koekemoer}, {Chang}, {McGrath},
  {H{\"a}ussler}, {Dekel}, {Behroozi}, {Fumagalli}, {Leja}, {Lundgren},
  {Maseda}, {Nelson}, {Wake}, {Patel}, {Labb{\'e}}, {Faber}, {Grogin}, \&
  {Kocevski}}]{vanderWel14}
{van der Wel}, A., {Franx}, M., {van Dokkum}, P.~G., {et~al.} 2014, \apj, 788,
  28

\bibitem[{{van Dokkum} \& {Franx}(1996)}]{vanDokkumFranx96}
{van Dokkum}, P.~G. \& {Franx}, M. 1996, \mnras, 281, 985

\bibitem[{{van Dokkum} {et~al.}(2015){van Dokkum}, {Nelson}, {Franx}, {Oesch},
  {Momcheva}, {Brammer}, {F{\"o}rster Schreiber}, {Skelton}, {Whitaker}, {van
  der Wel}, {Bezanson}, {Fumagalli}, {Illingworth}, {Kriek}, {Leja}, \&
  {Wuyts}}]{vanDokkum15}
{van Dokkum}, P.~G., {Nelson}, E.~J., {Franx}, M., {et~al.} 2015, \apj, 813, 23

\bibitem[{{Weinmann} {et~al.}(2009){Weinmann}, {Kauffmann}, {van den Bosch},
  {Pasquali}, {McIntosh}, {Mo}, {Yang}, \& {Guo}}]{Weinmann09}
{Weinmann}, S.~M., {Kauffmann}, G., {van den Bosch}, F.~C., {et~al.} 2009,
  \mnras, 394, 1213

\bibitem[{{Whitaker} {et~al.}(2011){Whitaker}, {Labb{\'e}}, {van Dokkum},
  {Brammer}, {Kriek}, {Marchesini}, {Quadri}, {Franx}, {Muzzin}, {Williams},
  {Bezanson}, {Illingworth}, {Lee}, {Lundgren}, {Nelson}, {Rudnick}, {Tal}, \&
  {Wake}}]{Whitaker11}
{Whitaker}, K.~E., {Labb{\'e}}, I., {van Dokkum}, P.~G., {et~al.} 2011, \apj,
  735, 86

\bibitem[{{Wilkinson} {et~al.}(2021){Wilkinson}, {Almaini}, {Wild}, {Maltby},
  {Hartley}, {Simpson}, \& {Rowlands}}]{Wilkinson21}
{Wilkinson}, A., {Almaini}, O., {Wild}, V., {et~al.} 2021, \mnras, 504, 4533

\bibitem[{{Williams} {et~al.}(2009){Williams}, {Quadri}, {Franx}, {van Dokkum},
  \& {Labb{\'e}}}]{Williams09}
{Williams}, R.~J., {Quadri}, R.~F., {Franx}, M., {van Dokkum}, P., \&
  {Labb{\'e}}, I. 2009, \apj, 691, 1879

\bibitem[{{Wuyts} {et~al.}(2007){Wuyts}, {Labb{\'e}}, {Franx}, {Rudnick}, {van
  Dokkum}, {Fazio}, {F{\"o}rster Schreiber}, {Huang}, {Moorwood}, {Rix},
  {R{\"o}ttgering}, \& {van der Werf}}]{Wuyts07}
{Wuyts}, S., {Labb{\'e}}, I., {Franx}, M., {et~al.} 2007, \apj, 655, 51

\bibitem[{{Wylezalek} {et~al.}(2013){Wylezalek}, {Galametz}, {Stern}, {Vernet},
  {De Breuck}, {Seymour}, {Brodwin}, {Eisenhardt}, {Gonzalez}, {Hatch},
  {Jarvis}, {Rettura}, {Stanford}, \& {Stevens}}]{Wylezalek13}
{Wylezalek}, D., {Galametz}, A., {Stern}, D., {et~al.} 2013, \apj, 769, 79

\bibitem[{{Wylezalek} {et~al.}(2014){Wylezalek}, {Vernet}, {De Breuck},
  {Stern}, {Brodwin}, {Galametz}, {Gonzalez}, {Jarvis}, {Hatch}, {Seymour}, \&
  {Stanford}}]{Wylezalek14}
{Wylezalek}, D., {Vernet}, J., {De Breuck}, C., {et~al.} 2014, \apj, 786, 17

\bibitem[{{Yang} {et~al.}(2021){Yang}, {Roberts-Borsani}, {Treu}, {Birrer},
  {Morishita}, \& {Brada{\v{c}}}}]{Yang21}
{Yang}, L., {Roberts-Borsani}, G., {Treu}, T., {et~al.} 2021, \mnras, 501, 1028

\bibitem[{{Yoon} {et~al.}(2017){Yoon}, {Im}, \& {Kim}}]{Yoon17}
{Yoon}, Y., {Im}, M., \& {Kim}, J.-W. 2017, \apj, 834, 73

\bibitem[{{Zanisi} {et~al.}(2021{\natexlab{a}}){Zanisi}, {Shankar}, {Bernardi},
  {Mei}, \& {Huertas-Company}}]{Zanisi21b}
{Zanisi}, L., {Shankar}, F., {Bernardi}, M., {Mei}, S., \& {Huertas-Company},
  M. 2021{\natexlab{a}}, \mnras, 505, L84

\bibitem[{{Zanisi} {et~al.}(2021{\natexlab{b}}){Zanisi}, {Shankar}, {Fu},
  {Rodriguez-Puebla}, {Avila-Reese}, {Faisst}, {Daddi}, {Boco}, {Lapi},
  {Giavalisco}, {Saracco}, {Buitrago}, {Huertas-Company}, {Puglisi}, \&
  {Dekel}}]{Zanisi21a}
{Zanisi}, L., {Shankar}, F., {Fu}, H., {et~al.} 2021{\natexlab{b}}, \mnras,
  505, 4555

\bibitem[{{Zanisi} {et~al.}(2020){Zanisi}, {Shankar}, {Lapi}, {Menci},
  {Bernardi}, {Duckworth}, {Huertas-Company}, {Grylls}, \&
  {Salucci}}]{Zanisi20}
{Zanisi}, L., {Shankar}, F., {Lapi}, A., {et~al.} 2020, \mnras, 492, 1671

\bibitem[{{Zhao} {et~al.}(2017){Zhao}, {Conselice}, {Arag{\'o}n-Salamanca},
  {Almaini}, {Hartley}, {Lani}, {Mortlock}, \& {Old}}]{Zhao17}
{Zhao}, D., {Conselice}, C.~J., {Arag{\'o}n-Salamanca}, A., {et~al.} 2017,
  \mnras, 464, 1393

\bibitem[{{Zoldan} {et~al.}(2019){Zoldan}, {De Lucia}, {Xie}, {Fontanot}, \&
  {Hirschmann}}]{Zoldan19}
{Zoldan}, A., {De Lucia}, G., {Xie}, L., {Fontanot}, F., \& {Hirschmann}, M.
  2019, \mnras, 487, 5649

\end{thebibliography}



\end{document}